\documentclass[iop]{emulateapj}

\usepackage{graphicx}
\usepackage{amsmath}
\usepackage{natbib}
\usepackage{mathptmx}
\usepackage{enumerate}
\usepackage{apjfonts}
\usepackage{times}
\usepackage[colorlinks=true,urlcolor=blue,citecolor=blue,linkcolor=blue]{hyperref}
\usepackage{aas_macros}

\bibpunct{(}{)}{;}{a}{}{,}

\usepackage[position=t,singlelinecheck=on,font={rm,md,up},font=normalsize]{subfig}
\makeatletter
\long\def\frontmatter@title@above{
  \vspace*{-\headsep}\vspace*{\headheight}
   {\sc The Astrophysical Journal}, 810:54 (12pp), 2015 September 1
  \par\vspace*{-\baselineskip}\vspace{12mm}
  }

\makeatother

\RequirePackage{color}

\newcommand\cites[1]{\citeauthor{#1}'s\ (\citeyear{#1})}

\shorttitle{Magnetic upflow events in the quiet-Sun photosphere. I. Observations}
\shortauthors{Jafarzadeh et al.}

\begin{document}

\title{Magnetic upflow events in the quiet-Sun photosphere. I. Observations}

\author{S.~Jafarzadeh\hyperlink{}{\altaffilmark{1}}}
\author{L.~Rouppe van der Voort\hyperlink{}{\altaffilmark{1}}}
\author{J.~de la Cruz Rodr\'iguez\hyperlink{}{\altaffilmark{2}}}

\affil{\altaffilmark{1}\hspace{0.2em}Institute of Theoretical Astrophysics, University of Oslo, P.O. Box 1029 Blindern, N-0315 Oslo, Norway; \href{mailto:shahin.jafarzadeh@astro.uio.no}{shahin.jafarzadeh@astro.uio.no}\\
\altaffilmark{2}\hspace{0.2em}Institute for Solar Physics, Department of Astronomy, Stockholm University, Albanova University Center, SE-10691 Stockholm, Sweden}

\begin{abstract}
Rapid magnetic upflows in the quiet-Sun photosphere were recently uncovered from both {\sc Sunrise}/IMaX and Hinode/SOT observations.
Here, we study magnetic upflow events (MUEs) from high-quality, high- (spatial, temporal, and spectral) resolution, and full Stokes observations in four photospheric magnetically sensitive Fe~{\sc i} lines centered at $5250.21$, $6173.34$, $6301.51$, and $6302.50$~\AA\ acquired with the Swedish Solar Telescope (SST)/CRISP. We detect MUEs by subtracting in-line Stokes $V$ signals from those in the far blue wing whose signal-to-noise ratio (S/N) $\geq7$.
We find a larger number of MUEs at any given time ($2.0\times10^{-2}$~arcsec$^{-2}$), larger by one to two orders of magnitude, than previously reported. The MUEs appear to fall into four classes presenting different shapes of Stokes $V$ profiles with (I) asymmetric double lobes, (II) single lobes, (III) double-humped (two same-polarity lobes), and (IV) three lobes (an extra blueshifted bump in addition to double lobes), of which less than half are single-lobed. We also find that MUEs are almost equally distributed in network and internetwork areas and they appear in the interior or at the edge of granules in both regions. Distributions of physical properties, except that of horizontal velocity, of the MUEs (namely, Stokes V signal, size, line-of-sight velocity, and lifetime) are almost identical for the different spectral lines in our data.
A bisector analysis of our spectrally resolved observations shows that these events host modest upflows and do not show a direct indication of the presence of supersonic upflows reported earlier.
Our findings reveal that the number, types (classes), and properties determined for MUEs can strongly depend on the detection techniques used and the properties of the employed data, namely, S/Ns, resolutions, and wavelengths.
\vspace{1mm}
\end{abstract}


\section{Introduction}
\label{sec:intro}

The solar photosphere is a highly dynamic environment where motions with a wide range of spatial and temporal scales are observed~\citep{Spruit1990}. Both magnetized and non-magnetized flows are found in the photosphere. The magnetized flows are related to or driven by the magnetic fields whose manifestation in the solar photosphere are on all spatial scales~\citep{Stenflo1989,Solanki1993,Solanki2001,deWijn2009}. Among those, observations of small-scale magnetic fields and their interactions with convective motions at high spatial and temporal resolutions have uncovered a number of interesting phenomena in the solar photosphere~\citep[e.g.,][]{BellotRubio2001,Shimizu2007b,Fischer2009,Martinez-Gonzalez2009,Zhang2009,Danilovic2010b,Steiner2010,Straus2010}. \cites{Shimizu2008} discovery of high-speed (supersonic; $\approx8-10$~km$\:$s$^{-1}$) downflows associated with the formation of small concentrated magnetic patches is one example of those small-scale magnetized flows to particularly mention here.

The magnetic fields in the solar atmosphere are inferred from measurements of polarization states of light through observations of magnetically sensitive lines (\citealt{Wittmann1974,Auer1977}; see also \citealt{Borrero2011} and references therein). The polarization states are represented by four Stokes parameters denoted by $I$ referring to the total intensity, $Q$ and $U$ describing linear polarizations, and $V$ characterizing circular polarization~\citep{Unno1956,Rachkovsky1962,Stenflo1971,Landi-DeglInnocenti12004}.

Recently, \citet{Borrero2010} claimed discovery of supersonic magnetic upflows from strong Stokes $V$ signals they found in the continuum position of the IMaX Fe~{\sc i} $5250.2$~\AA\ passband~\citep{Martinez-Pillet2011} aboard the {\sc Sunrise} balloon-borne solar observatory~\citep{Solanki2010}. Those large velocities were, however, inferred based on a single continuum position, from which the shape of Stokes $V$ profiles could not be sampled. 
The existence of these magnetic upflows was later confirmed by \citet{Martinez-Pillet2011b} who referred to them as photospheric jets. They made use of data from Hinode/SP~\citep{Tsuneta2008}, which have lower spatial resolution, but a higher spectral sampling compared to the data from {\sc Sunrise}. Using the IMaX data, \citet{RubiodaCosta2015} showed the number density of the events is independent of the heliocentric angle. They found a number density of $\approx7.0\times10^{-4}$~arcsec$^{-2}$, which is smaller than the values found by \citet{Borrero2010} and \citet{Martinez-Pillet2011b} ($\approx2.0\times10^{-3}$~arcsec$^{-2}$), but comparable to the number density reported by \citet{Borrero2013}.

These events were explained in terms of magnetic reconnection, where freshly emerged flux tubes within granules interact with pre-existing fields in intergranular areas~\citep{Borrero2013,QuinteroNoda2013}. \citet{QuinteroNoda2014} found that the extremely blueshifted events, observed by Hinode/SP, are often accompanied by redshifted Stokes $V$ signals. Together, they represent footpoints of magnetic $\Omega$-loops with flows along them. \citet{QuinteroNoda2014} argued that the siphon flow mechanism~\citep{Montesinos1993} was the only explanation for those flow motions along the arched magnetic flux tubes.

On the other hand, in the MHD simulations of \citet{Danilovic2015}, the rapid magnetic flows could solely be produced by flux emergence~\citep[e.g.,][]{Caligari1995,Lites1996,Orozco2008}. While magnetic reconnection could also take place in these simulations, it was not, necessarily, the cause of such events.

In the above studies, only single-lobed Stokes $V$ profiles were reported at the spatial locations of the rapid magnetic upflows, albeit \citet{Martinez-Pillet2011b} also noted observations of highly asymmetric Stokes $V$ profiles with multiple lobes (shifted with respect to the rest wavelength position). However, they did not analyze the profiles. We should note that few cases of supersonic magnetic upflows with double-peaked structures in the blue lobe of Stokes $V$ profiles (i.e., three-lobed Stokes $V$ profiles, where the blue lobes are double-peaked) were reported previously by \citet{Socas-Navarro2005}.

In this paper, we detect and study a wide range of magnetic Doppler shifted phenomena and in particular magnetic upflow events (MUEs) from observations with not only high spatial and temporal resolutions but also with highly sampled profiles (from the Swedish Solar Telescope (SST)/CRISP; Section~\ref{sec:data}). We also conduct our investigations on a time-series of images sampled in four different photospheric lines. In addition, we use a different method compared to that employed in previous studies with which a larger number of events (larger by one or two orders of magnitudes) are identified (Section~\ref{sec:detection}). Distributions of the physical and dynamical properties of the events are studied and their spatial distribution over network and internetwork areas is inspected (Section~\ref{sec:analysis}). We find four different classes of Stokes $V$ profiles corresponding to the MUEs. Finally, we discuss our results, compare them with those in the literature, and draw our conclusions in Section~\ref{sec:conclusions}.

\section{Data}
\label{sec:data}

In the present work, we use several high-quality, high-resolution data sets recorded at the quiet-Sun disk center by the CRisp Imaging SpectroPolarimeter (CRISP; \citealt{Scharmer2008}) mounted at the SST \citep{Scharmer2003}.
The combination of high overall transmission (with a minimum of optical surfaces) of the CRISP dual Fabry-P\'{e}rot Interferometer (FPI) and a high-cadence camera system provide a fast line sampling (with a typical effective sampling speed of 0.25~s per line position for non-polarimetric observations and 1--1.5~s per line position for the polarimetric data analyzed here) for a wide wavelength range of 5100--8600~\AA\ with numerous photospheric and chromospheric spectral diagnostics \citep{Scharmer2006}. In addition, the combination of high and low spectral resolution etalons at the CRISP FPI results in a high spectral resolution at a selected wavelength.
Polarization measurements are performed by two liquid crystal modulators in combination with a polarizing beam splitter which facilitates a large reduction of seeing-induced polarization crosstalk by simultaneously exposing two cameras.

The employed full Stokes ($I$, $Q$, $U$, and $V$) time series of images were acquired in four photospheric, magnetically sensitive, Fe~{\sc i} lines. Table~\ref{table:data} summarizes characteristics of these data sets along with the line parameters. To estimate the heights of formation of the four Fe~{\sc i} lines, we computed the line depression contribution functions using the RH radiative transfer code of \citet{Uitenbroek2001} in FALC, FALF, and FALP model atmospheres \citep{Fontenla1993,Fontenla2006}. These atmospheric models represent an averaged quiet-Sun area, bright regions of the quiet-Sun, and a typical plage region, respectively. We determined the contribution of the atmospheric heights to the radiation passing through the line core of our SST/CRISP Fe~{\sc i} lines (where the spectra were convolved with the transmission profiles of the filters) in non-LTE conditions. The formation heights given in Table~\ref{table:data} are the average values (weighted by the corresponding contribution functions) from the calculations in the FALC mode atmosphere. The average of the formation heights is slightly (less than 10 km) smaller in the other two atmospheric models compared to the FALC, for all the lines except for the 5250 line. The latter is smaller by $\approx50$~km in FALP compared to that obtained in the FALC model atmosphere.

The images have a scale of $\approx0.06$~arcsec~pixel$^{-1}$ covering an average field of view (FOV) of $52\times52$~arcsec$^2$ after restoration. The 6301/2 \AA\ data set also includes simultaneous observations in the Ca~{\sc ii}~H passband with a bandwidth of 1.0 \AA.\

\begin{table}[tp]
\caption{Overview of Lines and Data Sets from SST/CRISP}     
\label{table:data}
\setlength{\tabcolsep}{.5em}                       
\begin{tabular}{l c c c}    
\hline \hline
Quantity & Fe~{\sc i}~5250 \AA\ & Fe~{\sc i}~6173 \AA\ & Fe~{\sc i}~6301/6302 \AA\ \\
\tableline
   Line center & $5250.21$~\AA\ & $6173.34$~\AA\ & $6301.51$/$6302.50$~\AA\ \\   
   Land\'{e} factor\tablenotemark{a} & 3.0 & 2.5 & 1.67/2.5 \\
   $\log(g_{i}\,f_{ik})$\tablenotemark{b} & --4.938 & --2.880 & --0.718/--0.973 \\
\tableline
   Spectral resolution\tablenotemark{c} & $26.3$~m\AA\ & $50.7$~m\AA\ & $54.8$~m\AA\ \\
   Formation height\tablenotemark{d} & 214 km & 174 km & 178/125 km \\
\tableline
   Date of observations & 2014 Jun 23 & 2012 Jun 4 & 2011 Aug 6\\
   Duration & 13 minutes & 41 minutes & 47 minutes \\
   Cadence\tablenotemark{e} & 20 s & 14 s & 28 s \\
   Acquisition\tablenotemark{f} & 20 s & 14 s & 18 s\tablenotemark{g} \\
   No. of positions\tablenotemark{h} & 21 & 10 & 15/15 \\
   Sampling step & $25$~m\AA\ & $35$~m\AA\ \tablenotemark{i} & $44$~m\AA\ \\
   No. of exposures\tablenotemark{j} & 8 & 11 & 5 \\
   $\mu$\tablenotemark{k} & 1 & 1 & 1 \\
\tableline
\vspace{-1mm}
\end{tabular}
\hspace{1mm}
\vspace{0.7mm}
\textbf{Notes.}
\footnotetext[1]{ Magnetic sensitivity of the lines. From \citet{Socas-Navarro2004}.}
\footnotetext[2]{ A proxy for line strength between the lower ($i$) and upper ($k$) energy levels where $f_{ik}$ is the  oscillator strength and $g$ is the statistical weight. From \citet{Fabbian2012} and \citet{Kramida2014}.}
\footnotetext[3]{ FWHM of the CRISP transmission profile. From \citet{de-laCruz2015}.}
\footnotetext[4]{ Average heights of formation of the line cores for FALC model atmosphere, after convolution with the CRISP transmission profiles.}
\footnotetext[5]{ Time step in the time series of images.}
\footnotetext[6]{ Total time for acquisition of a full spectral scan.}
\footnotetext[7]{ This data set also includes observations of another line scan (Fe~{\sc i}~5576~\AA).}
\footnotetext[8]{ Number of wavelength positions.}
\footnotetext[9]{ The width of spectral sampling step between the second and eighth wavelength positions of this line. It is $70$~m\AA\ for the first and the ninth and $140$~m\AA\ for the last position.}
\footnotetext[10]{ Number of exposures per polarization state.}
\footnotetext[11]{ The cosine of the observing angle.}
\end{table}

\begin{figure}[!tp]
\vspace{0.3cm}
  \centerline{\includegraphics[width=8.5cm, trim = 0 0 0 0, clip]{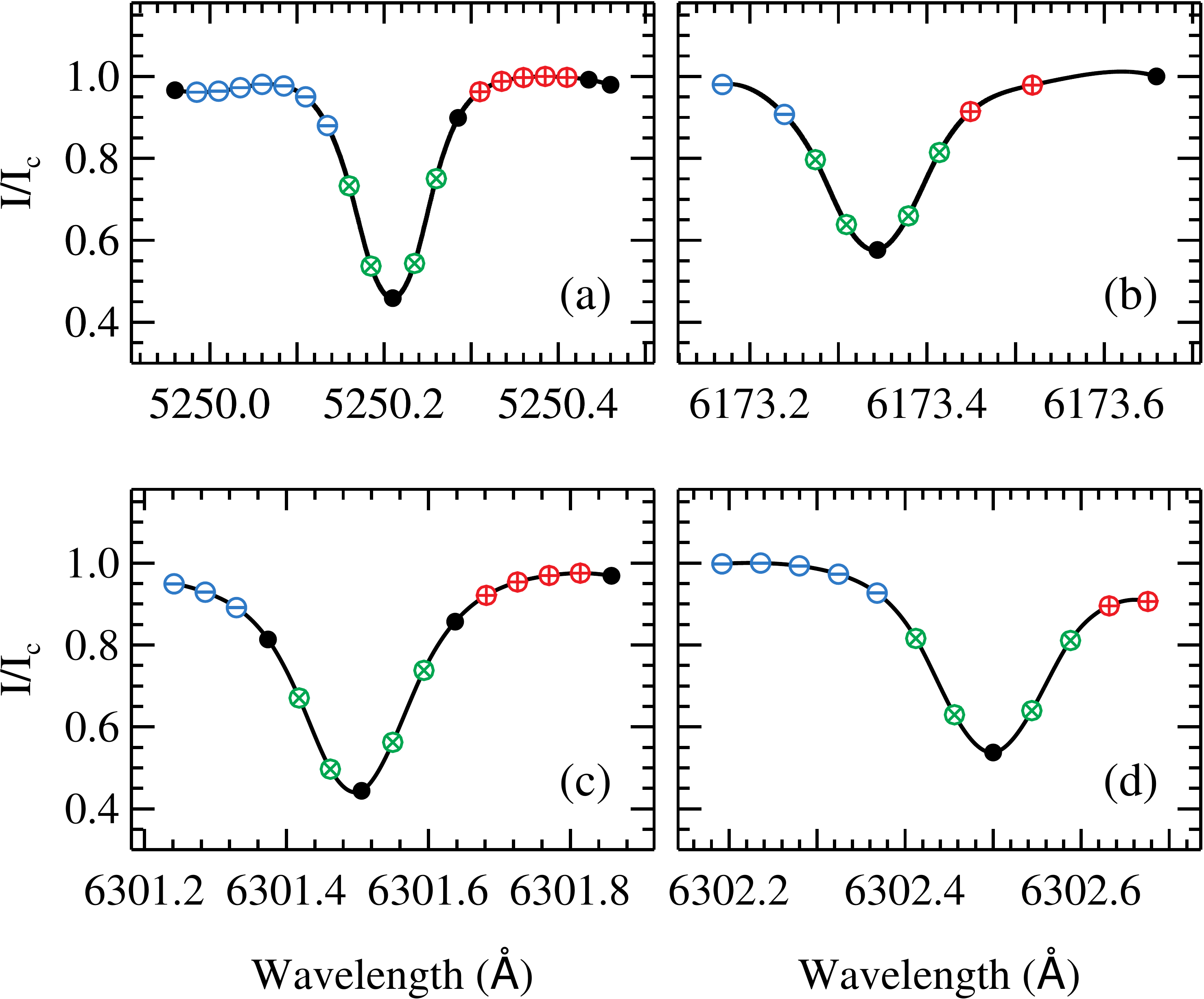}}
  \caption{Averaged Stokes $I$ profiles of the four photospheric Fe~{\sc i} lines centered at $5250.21$~\AA\ (a), $6173.34$~\AA\ (b), $6301.51$~\AA\ (c), and $6302.50$~\AA\ (d) whose Stokes $V$ profiles are analyzed in this paper. The circles indicate the sampled wavelength positions. The green, blue, and red circles (with cross, minus, and plus signs) mark those wavelength positions from which the averaged in-line, far-blue-wing, and far-red-wing Stokes $V$ images used for event detection are constructed.}
  \label{fig:stokesIps}
\end{figure}

All recorded images were processed using the CRISPRED reduction pipeline~\citep{de-laCruz2015} which also restores images from wavefront aberrations by means of Multi-Object Multi-Frame Blind Deconvolution (MOMFBD;~\citealt{vanNoort2005}). The pipeline also includes calibrations for field-/time-dependent instrumental polarization, inter-camera and temporal misalignments, as well as image field rotation, and corrections for reflectivity and cavity errors (induced in the CRISP FPI passband profiles).
We use multiple exposures per polarization state in the MOMFBD restoration which increases the signal-to-noise ratio (S/N) \citep{vanNoort2008}. Ultimately, the number of exposures per state is a trade-off between the necessity of high S/N dense spectral sampling, and the requirement to complete the line scan within the typical evolution timescale in the photosphere.
The adaptive optics system at the SST~\citep{Scharmer2003} as well as the MOMFBD post-processing approach result in images with high spatial resolution close to the diffraction limit of the 1 m telescope ($\lambda/D = 0\farcs1$ at 5250~\AA). 

Figure~\ref{fig:stokesIps} illustrates mean Stokes $I$ profiles of all the passbands (averaged over the entire FOV) along with their sampled wavelength positions marked with circles. For the purpose of event detection, we combine several line positions to represent different parts of the profile at higher signal-to-noise. Different colors/shapes of the circles indicate the selected wavelength positions from which averaged Stoke $V$ signals are combined in the far blue wing ($V^{-}_\mathrm{wing}$; blue circles), far red wing ($V^{+}_\mathrm{wing}$; red circles), and in-line positions close to the line-core ($V_\mathrm{line}$; green circles). To avoid cancellation, the signs of the wavelength points on the red side of the line core are reversed prior to forming the in-line images.
These averaged Stokes $V$ images have a larger S/N compared to the individual wavelength points (the noise levels are reduced by the square root of the number of wavelength positions with which the averaged images are formed). Hence, their absolute values are being used for detection of the magnetic Doppler-shifted events (see Section~\ref{sec:detection}). Thus, three sets of images, from each Stokes $V$ profile, are formed as

\begin{equation}
	\left |V_\mathrm{line}   \right |= \frac{1}{n_{1}+n_{2}}\left | \sum_{i}^{n_{1}} V(\lambda _{i}) - \sum_{i}^{n_{2}} V(\lambda _{i}) \right |\,,
	\label{equ:vline}
\end{equation}

\begin{equation}
	\left |V^{-}_\mathrm{wing}   \right |= \frac{1}{n^{-}}\left | \sum_{i^{-}}^{n^{-}} V(\lambda _{i^{-}}) \right |\,,
	\label{equ:vwingb}
\end{equation}

\begin{equation}
	\left |V^{+}_\mathrm{wing}   \right |= \frac{1}{n^{+}}\left | \sum_{i^{+}}^{n^{+}} V(\lambda _{i^{+}}) \right |\,,
	\label{equ:vwingr}
\end{equation}

\noindent
where the indices $i$, $i^{-}$, and $i^{+}$ represent the scanning wavelength positions indicated in Figure~\ref{fig:stokesIps} by the green circles with cross signs, blue circles with minus signs, and red circles with plus signs, respectively, running from the bluest to reddest points for the number of corresponding marked circles ($n_{1}$, $n_{2}$, $n^{-}$, and $n^{+}$).

The Stokes parameters have a $1\sigma$ noise level of $\approx1.6\times10^{-3}~I_{c}$ per wavelength, after post-processing. The noise level was computed as the standard deviation at a far continuum position of the Stokes profiles where almost no signal is apparent. We note that the $1\sigma$ noise level of $V^{-}_\mathrm{wing}$, $V^{+}_\mathrm{wing}$ and $V_\mathrm{line}$ images are reduced down to $(6.0-11.0)\times10^{-4}~I_{c}$ depending on the number of wavelength positions used to form the averaged images (see Figure~\ref{fig:stokesIps}).

For the sake of simplicity and of consistency, we will show all examples (images and profiles) in the following for those recorded in the Fe~{\sc i} 6301/6302 lines. The same analyses are employed for all four spectral lines (unless otherwise stated). The number of detected events from the different passbands (and some of their physical properties) are, however, different (see Section~\ref{sec:analysis}).

\section{Event detection}
\label{sec:detection}

We aim to study Doppler-shifted (in particular blueshifted) events associated with the magnetic fields. To this end, we want to detect Stokes $V$ signals that are largely shifted toward the blue or toward the red wavelengths, i.e., where the Stokes $V$ signal in a pixel is significantly larger in the far wings compared to the in-line signal of the same pixel. In the following, we refer to these features as magnetic upflow events (MUEs) and magnetic downflow events (MDEs). We will, however, exclude those events that are smaller than the spatial resolution of our observations (i.e., $\approx10$~pixels in area). In addition, in order to secure detections of true signals, we exclude pixels whose far-wings Stokes $V$ signals are smaller than their corresponding $7\sigma$ noise levels (i.e., $0.6\%~I_{c}$ on average; in the range of $0.4\%-0.8\%~I_{c}$ for the different sampled lines).

For the event detection, we propose an approach that is based on subtraction of $|V_\mathrm{line}|$ from $|V^{\mp}_\mathrm{wing}|$. To compare our method with that employed in the earlier works by, e.g., \citealt{Borrero2013}, we also apply a technique that is defined as the quotient of the far wings and in-line signals.

\subsection{Difference Method}
\label{subsec:subtract}

Positive values of the difference between absolute values of far-wings Stokes $V$ signals (far blue wing, $|V^{-}_\mathrm{wing}|$, and/or far red wing ,$|V^{+}_\mathrm{wing}|$) and those of in-line ones ($|V_\mathrm{line}|$) represent Stoke $V$ signals that are largely shifted toward blue or red wavelengths corresponding to magnetic upflows and magnetic downflows, respectively. This builds the base for our event detection approach. We should, however, make sure that the real signal is properly distinguished from the noise prior to the subtraction. Thus, we set the $7\sigma$ noise level threshold on the $|V^{\mp}_\mathrm{wing}|$ images. The definitions of magnetic upflows/downflows are then denoted as

\begin{equation}
	\mathrm{MUE} :\overset{\underset{\mathrm{def}}{}}{=} \;\;  \left |V^{-}_\mathrm{wing}   \right |_\mathrm{sig} - \left |V_\mathrm{line}   \right |> 0 \,,
	\label{equ:mues}
\end{equation}

\begin{equation}
	\mathrm{MDE} :\overset{\underset{\mathrm{def}}{}}{=} \;\;  \left |V^{+}_\mathrm{wing}   \right |_\mathrm{sig} - \left |V_\mathrm{line}   \right |> 0 \,,
	\label{equ:mdes}
\end{equation}

\noindent
where

\begin{equation}
	\left |V^{\mp}_\mathrm{wing}   \right |_\mathrm{sig} :\overset{\underset{\mathrm{def}}{}}{=} \;\;  \left |V^{\mp}_\mathrm{wing}   \right |\geqslant 7\sigma_\mathrm{noise}\,.
	\label{equ:fwsig}
\end{equation}

As noted earlier, we only keep those events whose areas are larger than 10 pixels. Finally, we do a visual inspection of all individual frames to exclude possible false (or unclear) detected events. The latter could be a result of, e.g., image artifacts (fringes) coinciding at the location of the candidate MUEs/MDEs.
We note that the absolute values of MUEs/MDEs signals are only used for detection purposes. Both events have been found to be of either polarity.

\subsection{Quotient Method}
\label{subsec:quotient}

For comparison reasons, we also detect MUEs (from the same data sets as in Section~\ref{subsec:subtract}) by employing a similar approach as described by \citet{Borrero2013}. In this method, the absolute value of the quotient of far-blue-wing and in-line Stokes $V$ signals is determined and the MUEs are defined as $|V^{-}_\mathrm{wing}/V_\mathrm{line}|\gg1$.

For a direct comparison with our difference approach, we implement the quotient method after masking $|V^{-}_\mathrm{wing}|$ with $7\sigma_\mathrm{noise}$ of the employed data sets, and similarly exclude MUEs smaller than 10 pixels. Also, we redo the quotient method when $|V^{-}_\mathrm{wing}/V_\mathrm{line}|>4$ and without the earlier criteria, i.e., same as employed by \citet{Borrero2013}. See Section~\ref{sec:conclusions} for comparisons and discussions.

\subsection{A Case Study}
\label{subsec:casestudy}

\begin{figure*}[!]
\centering
  \subfloat{%
    \includegraphics[width=18cm, trim = 0 0 0 0, clip]{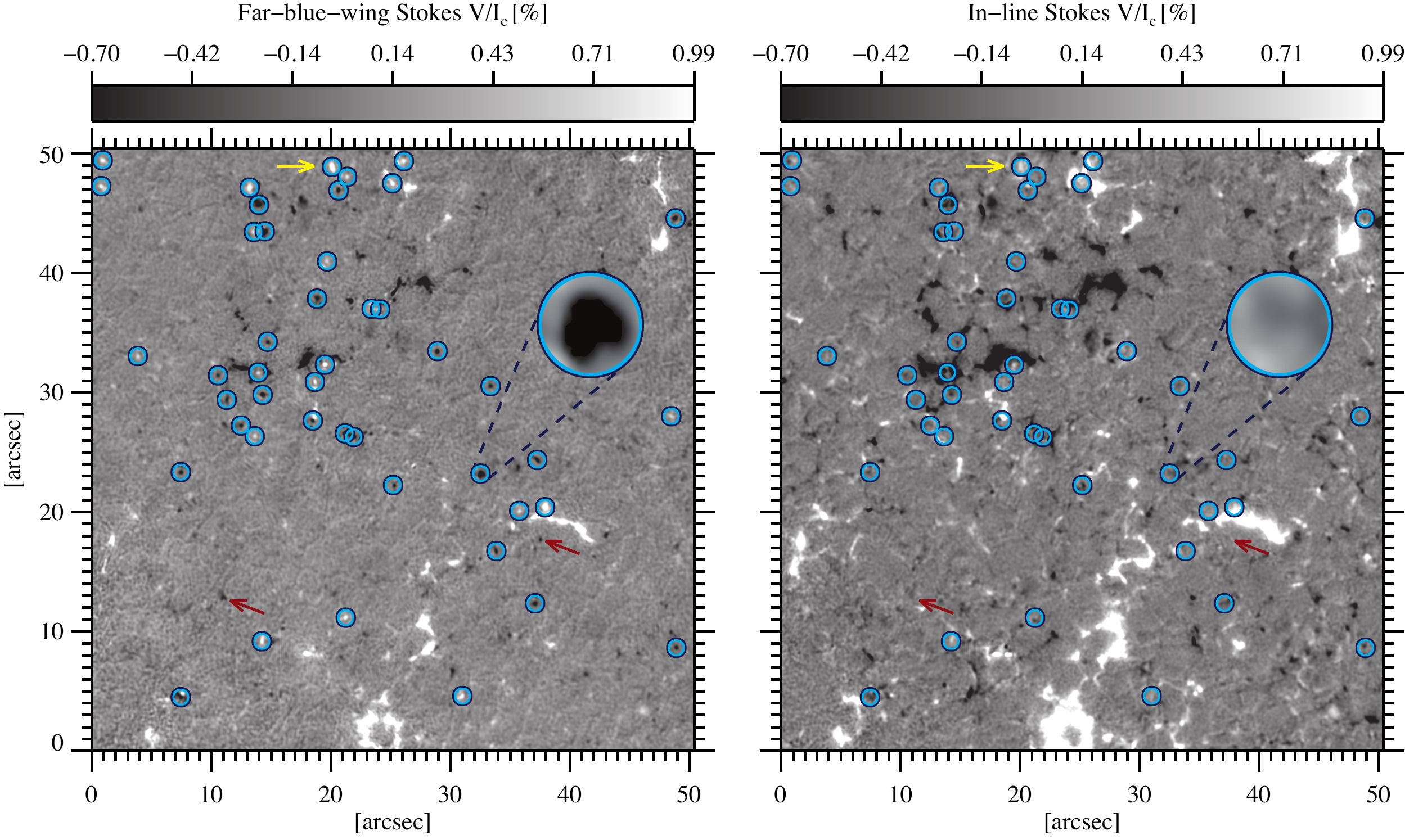}}\\
  \subfloat{%
    \includegraphics[width=18cm, trim = 0 0 0 0, clip]{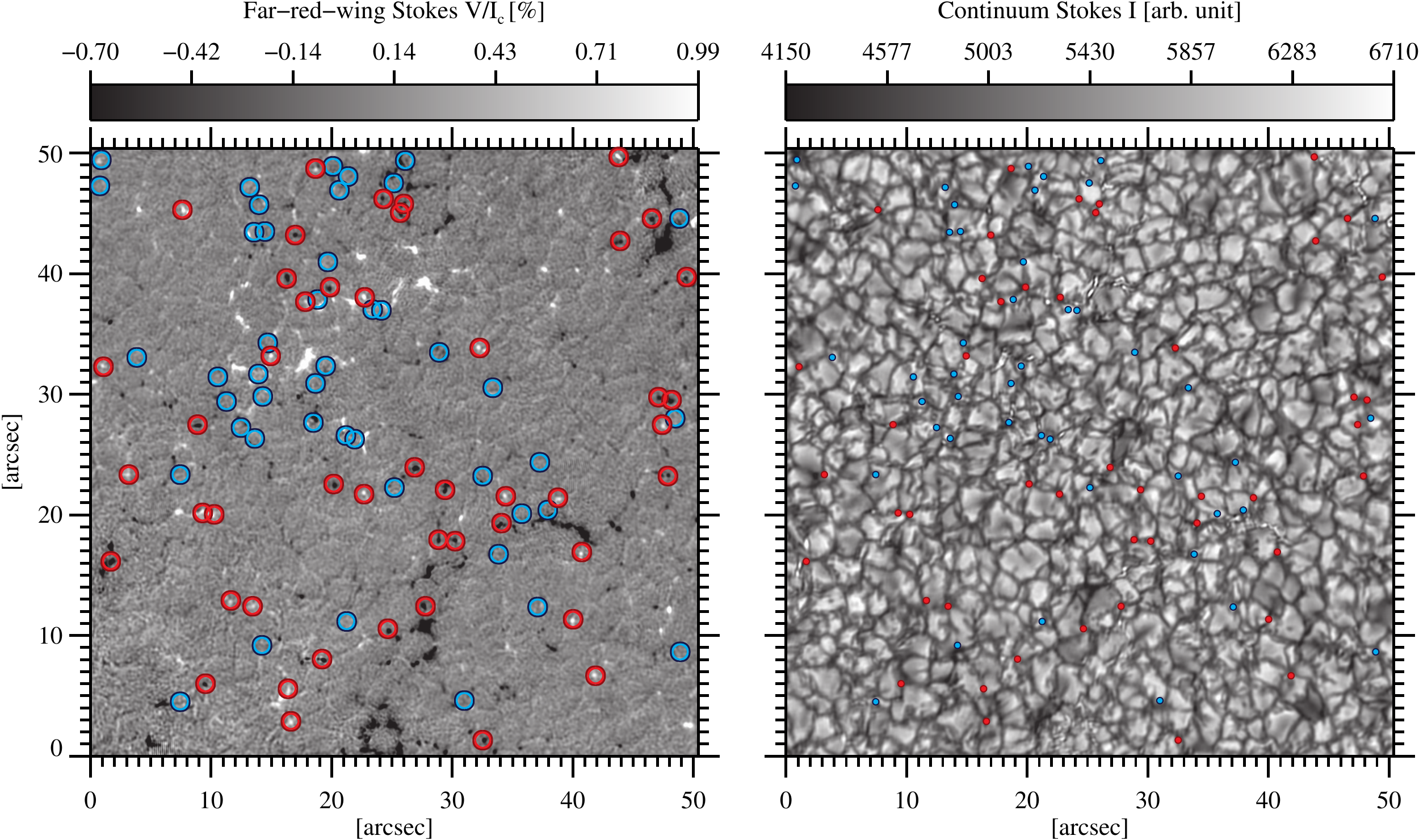}}\\
  \caption{Examples of far-blue-wing Stokes $V$ (top left), in-line Stokes $V$ (top right), far-red-wing Stokes $V$ (bottom left) and Stokes $I$ continuum (bottom right) images from the Fe~{\sc i} $6302$~\AA\ line. The Stokes $V$ images are saturated to the same scale. The blue and red circles mark detected magnetic upflow/downflow events (MUEs/MDEs), respectively. Both events are of either polarities and absolute values of their signals are only used for detection purposes.
On the upper panels: one of the MUEs is magnified, the red arrows indicate two examples of apparent MUE candidates that do not meet signal threshold criterion, and the yellow arrows mark an example where the MUE visually shows strong signals in both images.}
  \label{fig:MUEs_MDEs}
\end{figure*}

Illustrated in Figure~\ref{fig:MUEs_MDEs} are examples of the far-blue-wing (top left), far-red-wing (bottom left), and in-line (top right) Stokes $V$ images along with an image of Stokes $I$ continuum (bottom right) recorded in a scan of the $6302$~\AA\ line (see Section~\ref{subsec:subtract} for the definitions). For better visibility and for further visual comparisons, all the Stokes $V$ images are saturated to the same scale. The actual signals of the $V^{-}_\mathrm{wing}$, $V^{+}_\mathrm{wing}$, and $V_\mathrm{line}$ images in Figure~\ref{fig:MUEs_MDEs} are in the range of (-2.4\%,3.3\%), (-3.6\%,3.0\%), and (-7.3\%,8.5\%), relative to the Stokes $I$ continuum ($I_{c}$), respectively. The blue circles in Figure~\ref{fig:MUEs_MDEs} include the detected MUEs from our difference method. We remind the reader that the events may have either polarities and absolute values of their signals are only used to facilitate their detection.
One of the MUEs is magnified within its circle (all open circles have an inner diameter of $\approx1$~arcsec).
A visual comparison of the events, indicated by the blue circles, shows a significantly larger Stokes $V$ signal in the far-blue-wing image compared to the in-line image. This is not, however, clearly seen for some of the events. There exist pixels whose co-spatial locations may visually show strong signals in both $V^{-}_\mathrm{wing}$ and $V_\mathrm{line}$ images, while the signals are actually much larger in the $V^{-}_\mathrm{wing}$ than in the $V_\mathrm{line}$. The yellow arrows on the upper panels mark one of the detected events of the latter case. The red arrows indicate two examples of events that look like detected MUEs (with $V^{-}_\mathrm{wing}>V_\mathrm{line}$) but are not because they do not meet the $|V^{-}_\mathrm{wing}|\geq7\sigma_\mathrm{noise}$ criterion.
We note again that in addition to the signal threshold of $7\sigma_\mathrm{noise}$ for the $|V^{-}_\mathrm{wing}|$, we only consider events whose sizes are larger than 10 pixels (i.e., the same order of the resolution limit of the observations). Therefore, all relatively small MUEs, which are mostly found in internetwork areas, are excluded in this paper. These relatively small events may not be clearly seen by visual inspections in Figure~\ref{fig:MUEs_MDEs}. 

Similarly, we also identify MDEs from far-red-wing Stokes $V$ images and investigate correlations between their spatial locations with those of MUEs detected from the same scan. 
The bottom-left panel of Figure~\ref{fig:MUEs_MDEs} displays both MUEs and MDEs, marked with blue and red circles, respectively. The MDEs are derived from the far red wing of the same Stokes $V$ profile from which the MUEs are derived. No clear correlation between the spatial locations of MUEs and MDEs is observed.

A visual inspection of the spatial locations of both MUEs and MDEs on Stokes $I$ continuum images revealed that almost all MUEs are located on the granules, close to their edges, whereas MDEs are preferably observed on intergranular areas. This is, however, not so clear for a few of each of the two kinds of events whose relatively large areas overlap with the edges of granules (i.e., apparently are located on the boundaries between granules and intergranules). The bottom right panel of Figure~\ref{fig:MUEs_MDEs} represents the Stokes $I$ continuum image corresponding to the Stokes $V$ images. Positions of the detected MUEs and MDEs are depicted by small blue and red filled circles, respectively (for the sake of simplicity, sizes of the circles do not represent the actual sizes of the events).

A total of 45 MUEs and 47 MDEs are found (using our difference approach) in the example scan represented in Figure~\ref{fig:MUEs_MDEs}. We note that our detection algorithm prevents false detection (due to, e.g., spatially unresolved and/or relatively weak Stokes $V$ signals affected by noise) at the expense of a smaller number of detected MUEs/MDEs.

A much smaller number of MUEs (i.e., 17) are detected in the same scan by employing the quotient method (when a $7\sigma_\mathrm{noise}$ threshold of $|V^{-}_\mathrm{wing}|$ is applied prior to the detection).

\section{Analysis and statistics}
\label{sec:analysis}

We employ the same procedure as described in Section~\ref{subsec:casestudy} (i.e., the same detection algorithm and criteria) to all of the image sequences of the four observed passbands. The detected MUEs are then tracked in the time series of images and their temporal properties, namely horizontal velocity and lifetime, are determined. We use the same code as described in \citet{Jafarzadeh2013a} for tracking the MUEs. Sizes and amplitudes of Stokes $V$ signals of all detected MUEs are also calculated. For the latter, we take maximum Stokes $V$ signals of all individual pixels within the detected MUEs. We also investigate how the MUEs are distributed over network and internetwork areas.
In addition, we classify Stokes $V$ profiles of the detected MUEs. The MUEs are also detected from the quotient method (Section~\ref{subsec:quotient}), but only for comparison reasons (i.e., to compare the number densities of the events from the two approaches). We do not study MDEs in detail since that is beyond the scope of this paper. Their detections are only used to investigate possible correlations between their spatial locations and those of MUEs.

To avoid the effect of particular limitations of a particular data set on the final quantities (resulting in probable misinterpretations), the above analyses are carried out on the data sets whose temporal and/or spectral samplings are appropriate for certain parameters. Thus, the dynamical properties (lifetime and horizontal velocity) of the MUEs are only determined for our relatively long time series of images recorded in the 6173, 6301, and 6302~\AA\ lines. The classification of Stokes $V$ profiles are only done for the lines whose far blue wings are highly (spectrally) sampled, i.e., the 5250 and 6301/2~\AA\ lines. The other physical parameters (Stokes $V$ signal, size, and line-of-sight velocity) are measured for all the datasets. 
We note that the physical and dynamical parameters of the MUEs observed in the 6301 and 6302~\AA\ pair lines are determined independently for the individual lines (with different characteristics, e.g., heights of formation; see Table~\ref{table:data}). However, simultaneously observed MUEs are counted only once in the number statistics.
The region (network/internetwork) association of the MUEs is primarily investigated for the latter data set (6301 and 6302~\AA\ pair lines), from which simultaneous observations in the Ca~{\sc ii}~H passband (used to identify network areas) are also available. This is, however, also inspected for all the lines in our data sets, but from time-averaged Stokes $V$ images.

\subsection{Physical and Dynamical Properties}
\label{subsec:properties}

We detected a total number of 26,517 individual MUEs (covering 722,798 pixels in total) from all individual frames of the image sequences recorded in the four photospheric Fe~{\sc i} passbands (see Section~\ref{sec:data} for a description of the data sets). It comprises 5150 independent MUEs (when each MUE is counted once during its entire lifetime). We note that MUEs detected in 6301/2 \AA\ lines were counted once since the same events were observed in the pair lines.
We found the same order of magnitude but different rates of occurrence of the MUEs detected in the different passbands: $1.8\times10^{-2}$, $3.9\times10^{-2}$, and $1.5\times10^{-2}$~arcsec$^{-2}$ at any given time from images sampled in 5250, 6173, and 6301/2 \AA\ lines, respectively. It obviously makes the number of detected events in the 6173~\AA\ images about twice as larger as those found in the other two passbands.

\begin{figure}[!b]
  \centerline{\includegraphics[width=8.5cm, trim = 0 0 0 0, clip]{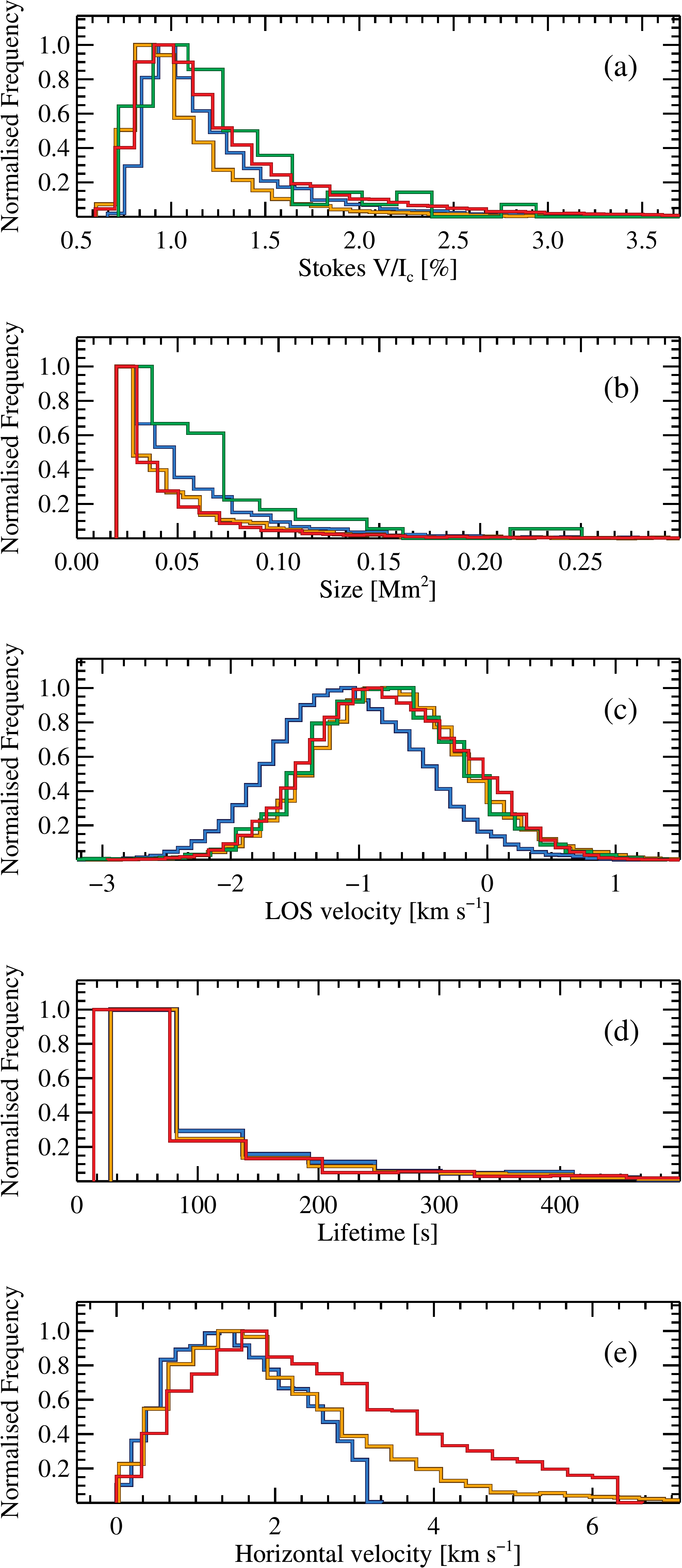}}
  \caption{Distributions of Stokes $V$ signals, normalized to Stokes $I$ continuum (a), lifetimes (b), line-of-sight (LOS) velocities (c), sizes (d), and horizontal velocities (e) of the detected magnetic upflow events (MUEs). The LOS velocities were constructed from bisectors at the wings of the Stokes $I$ profiles. The green, red, yellow, and blue histograms represent the parameters of MUEs observed in the Fe~{\sc i}~$5250$~\AA,\ Fe~{\sc i}~$6173$~\AA,\ Fe~{\sc i}~$6301$~\AA,\ and Fe~{\sc i}~$6302$~\AA\ passbands, respectively, where applicable (see the main text). The histograms are normalized to their maximum densities.}
  \label{fig:histograms}
\end{figure}

The area of individual MUEs (i.e., the number of pixels that each event contains) and the maximum Stokes $V$ signal of each individual pixel within the MUEs were determined.
We also measured the LOS velocity at each pixel (within the MUEs) from bisector positions. The latter are mid-points of horizontal segments between the two wings of Stokes $I$ profiles. These were measured at different line depths between intensity levels of 0\% (line core) and 80\% (close to continuum level). The difference between the average of the bisector positions over the different line depths of a profile and a reference bisector position results in a mean line shift that is translated to the Doppler velocity of that profile (pixel). The reference bisector position (or the reference wavelength) was computed by averaging mean bisector positions of all profiles over the entire FOV.

\begin{table*}[!tp]
\hspace{80mm} 
\caption{Summary of Properties of Magnetic Upflow Events (MUEs) Detected in Different Time Series of Images Sampled in Different Photospheric Passbands}     
\label{table:stats}
\centering
\setlength{\tabcolsep}{0.99em}                  
\begin{tabular}{l l c c c c}      
\hline \hline              
Parameter & Quantity & 5250 \AA\ & 6173 \AA\ & 6301 \AA\ & 6302 \AA\ \\   
\hline                      
   Stokes $V/I_{c}$ [\%] & Range & 0.7 , 3.8 & 0.6 , 4.9 & 0.6 , 3.6 & 0.7 , 4.4 \\   
   			   & Mean & 1.3 & 1.3 & 1.1 & 1.3 \\
\hline
   Size [Mm$^2$] & Range & 0.02 , 0.25 & 0.02 , 0.45 & 0.02 , 0.36 & 0.02 , 0.40  \\
   			   & Mean & 0.06 & 0.05 & 0.05 & 0.06 \\ 
\hline
   LOS velocity \tablenotemark{a} [km$\:$s$^{-1}$] & Range & -5.8 , 1.2 & -3.0 , 1.4 & -3.1 , 2.5 & -3.3 , 1.2 \\
   			   & Mean & -0.8 & -0.8 & -0.7 & -1.1 \\ 
\hline
   Lifetime [s] & Range & - & 14 , 980 & 28 , 1120 & 28 , 1120 \\
   			   & Mean & - & 113 & 114 & 113 \\
\hline
   Horizontal velocity [km$\:$s$^{-1}$] & Range & - & 0.0 , 6.3 & 0.0 , 7.8 & 0.0 , 3.2 \\
    & Mean & - & 2.6 & 2.1 & 1.5 \\ 
\hline  
   Number of MUEs & individual & 2958 & 19268 & 4291 & 4291 \\
    & independent & 406 & 3730 & 1014 & 1014 \\
\hline  
   Rate of occurrence [arcsec$^{-2}$] &  & $1.8\times10^{-2}$ & $3.9\times10^{-2}$ & $1.5\times10^{-2}$ & $1.5\times10^{-2}$ \\
\tableline                          
\end{tabular}
\footnotetext[1]{ From bisectors at the wings of Stokes $I$ profiles.}
\end{table*}

We further tracked the MUEs in the image sequences and measured their lifetimes and horizontal velocities. The latter were obtained as a result of dividing their frame-to-frame displacements by the time differences between the two consecutive frames.

Distributions of the Stokes $V$ signals (normalized to Stokes $I$ continuum), sizes, bisector LOS velocities, lifetimes, and horizontal velocities of the MUEs are represented in Figure~\ref{fig:histograms}(a)-(e), respectively. The green, red, yellow, and blue lines represent the histograms from 5250, 6173, 6301, and $6302$~\AA\ lines, respectively, where applicable.

Table~\ref{table:stats} summarizes ranges and mean values of all parameters determined from the different passbands along with the number of detected MUEs in each data set. We note that the lower and/or upper limits of the parameters presented in Table~\ref{table:stats} often lie outside the range of the histograms shown in Figure~\ref{fig:histograms}. All quantities have somewhat different but overlapping histograms obtained from observations in different wavelengths. Interestingly, the mean values of their Stokes $V$ signals, sizes, LOS velocities, and lifetimes (averaged over all MUEs) are almost identical to different passbands. The upflows are, however, slightly larger for the MUEs observed in 6302 \AA\ than those in the other three lines. About 1\% of the 5250 \AA\ MUEs tend to rather have rapid upflows with speeds larger than 5~km$\:$s$^{-1}$.

There are some differences between the mean values (and also upper ranges) of horizontal velocities of the MUEs. In particular, the MUEs observed in the $6173$~\AA\ passband move, on average, faster in the horizontal plane compared to those found in the $6301/2$~\AA\ images. While the $6301$ MUEs have also a lager proper motion, on average, than those observed in the $6302$ images, the horizontal velocity of a number of MUEs detected in the $6173$~\AA\ and $6301$~\AA\ datasets reach to a speed comparable to that of sound in the solar photosphere.

The lower limits of the sizes and lifetimes are limited to our size criterion of 10 pixels and the cadence of observations, respectively. For the latter case, the lifetime of a MUE was assumed equal to the cadence of the corresponding dataset when it was observed in only one frame.

\subsection{Region association}
\label{subsec:region}

\begin{figure}[tp!]
  \centerline{\includegraphics[width=8.6cm, trim = 0 0 0 0, clip]{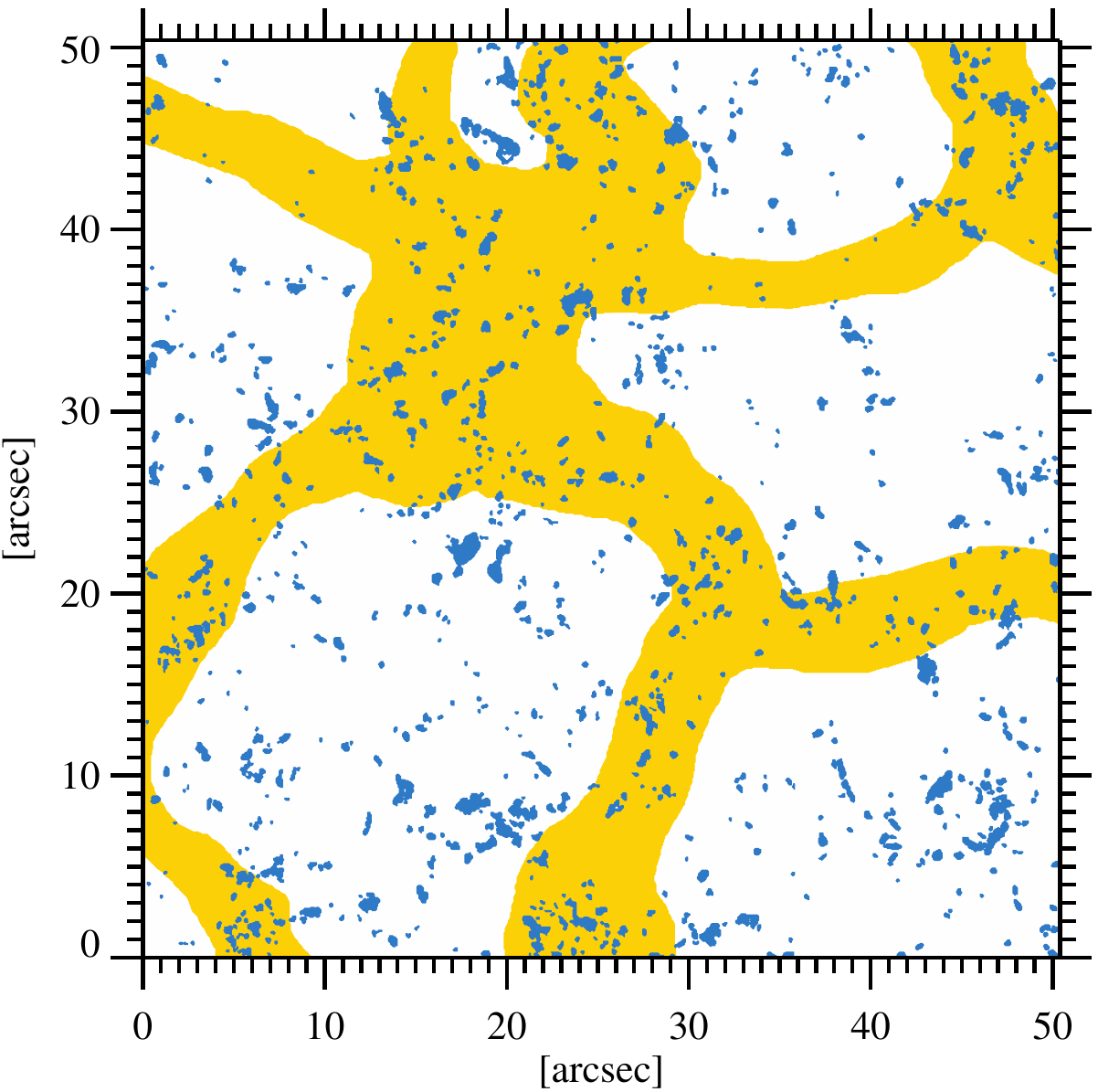}}
  \caption{All detected magnetic upflow events (MUEs) over the 6301/2 \AA\ image sequence of 47 minutes. The yellow and white areas represent network and internetwork regions, respectively, determined from the temporal average of simultaneously recorded Ca~{\sc ii}~H images.}
  \label{fig:allMUEs}
\end{figure}

The image sequences of 6301/2~\AA\ pair lines include simultaneous observations of Ca~{\sc ii}~H filtergrams. Taking into account the fact that regions with enhanced brightness in Ca~{\sc ii}~H images represent network patches, we create a mask from their time average from which the network and internetwork areas can also be distinguished in the (spatially and temporally) co-aligned 6301/2~\AA\ images. Using this mask, we investigate the distribution of MUEs over the network and internetwork regions. A similar comparison is feasible with the temporal average of Stokes $V$ maps where concentrations of the magnetic fields representing the network areas are observed. Figure~\ref{fig:allMUEs} includes all detected MUEs from the entire time-series of images observed in $6301/2$~\AA\ passband, overplotted in one frame. The yellow filled contour indicates the network areas. Hence, the white regions represent the internetwork. It turned out that about $58\%$ of all selected MUEs are located in internetwork regions, while the rest ($42\%$) are observed in network areas. We note that although these fractions are biased by the manual creation of the mask, it reveals that the MUEs are approximately equally distributed in the network and internetwork areas. Similar results were obtained for the MUEs observed in all three lines where network/internetwork masks were created from their corresponding (saturated) mean Stokes $V$ images (averaged over their entire image sequences).

\subsection{Classification of Stokes $V$ profiles}
\label{subsec:classify}

The Stokes $V$ profiles in the MUEs (from $5250$ and $6301/2$~\AA\ datasets) represent various ``anomalous'' shapes.
We carry out a classification of these profiles similar to a strategy described by \citet{Sigwarth2001}.
The $V$ profiles in the $6173$~\AA\ MUEs were poorly sampled in the far blue wing (resulting in unclear shapes), thus they are excluded from this classification.

Regardless of any assumptions for their origins and of the fact that the different shapes of Stokes $V$ profiles may correspond to different types of events, we classify them solely based on their deviations from a normal (double-lobed, antisymmetric) Stokes $V$ profile. This is conducted by visual inspections of the mean profiles (averaged over all pixels of the individual events) according to their appearances. The ratio between amplitudes of blue and red lobes are used to subdivide asymmetric double-lobed Stokes $V$ profiles (see below). Only ``lobes'' lying above the corresponding $3\sigma$ noise levels are considered as real signals, hence are included in the classification.

We find four classes of Stokes $V$ profiles as follows (of which one includes two sub-classes):
\begin{enumerate}[(I)]
\item \textit{Asymmetric double-lobed profiles} have two asymmetric lobes with opposite polarities. Their zero-crossing points are shifted toward the blue wavelengths in respect to the reference position (i.e., the rest wavelength of line core of corresponding lines; see Table~\ref{table:data}).
\begin{enumerate}
	\item \textit{Profiles with small asymmetries} have a ratio between their blue and red lobes that is smaller than 50\%.
	\item \textit{Profiles with large asymmetries} have one lobe with an amplitude larger than the other by a factor of at least two. 
\end{enumerate}
\item \textit{Asymmetric single-lobed profiles} have only one lobe whose amplitude lies above the $3\sigma$ noise level.
\item \textit{Double-humped profiles} have two lobes of the same polarity.
\item \textit{Double-lobed profiles with an extra blue-shifted bump} consist an extra, highly blue shifted bump that is of opposite polarity to the main double-lobed profile. These profiles look similar to those of typical Stokes $Q$ and $U$, shifted toward blue wavelengths.
\end{enumerate}

For the sake of consistency, we used same terminology as in the literature~\citep[e.g.,][]{Grossmann-Doerth2000,Sigwarth2001,Borrero2013} where possible.

\begin{figure*}[!t]
\centering
\renewcommand{\thesubfigure}{(Ia)}
  \subfloat[]{%
    \includegraphics[width=8.6cm, trim = 0 0 0 0, clip]{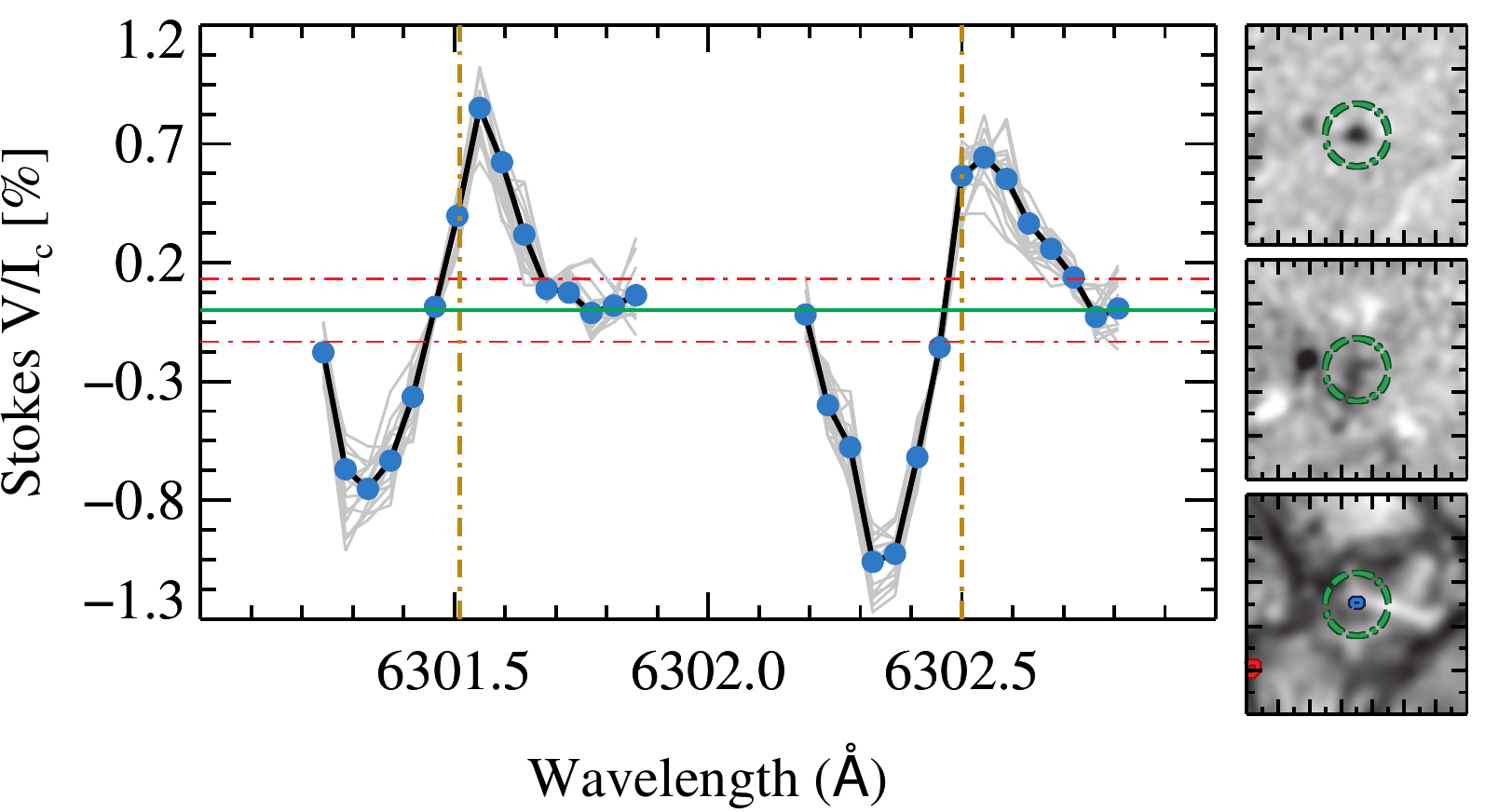}}\hfill
\renewcommand{\thesubfigure}{(Ib)}
  \subfloat[]{%
    \includegraphics[width=8.6cm, trim = 0 0 0 0, clip]{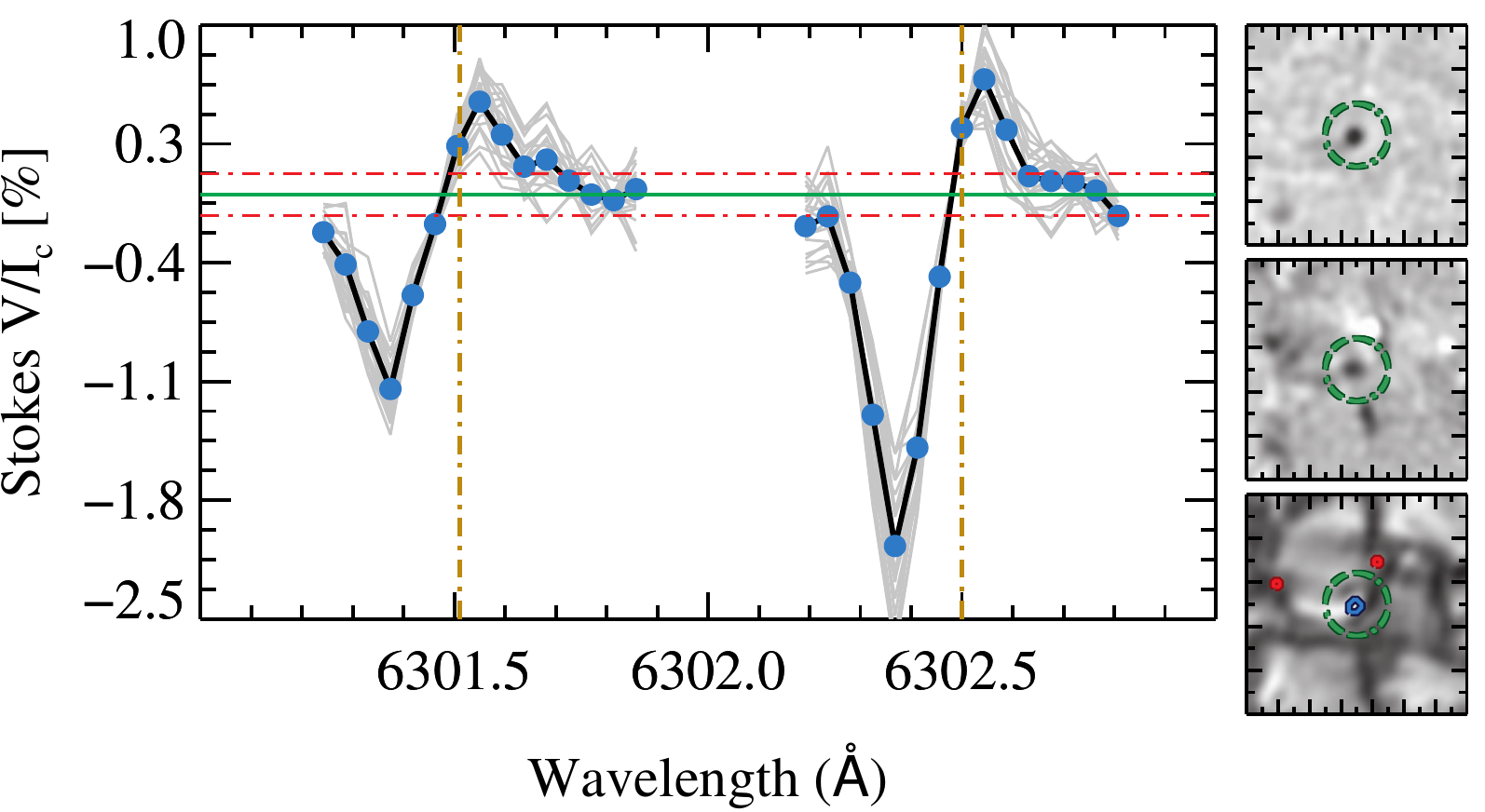}}\\
\renewcommand{\thesubfigure}{(II)}
  \subfloat[]{%
    \includegraphics[width=8.6cm, trim = 0 0 0 0, clip]{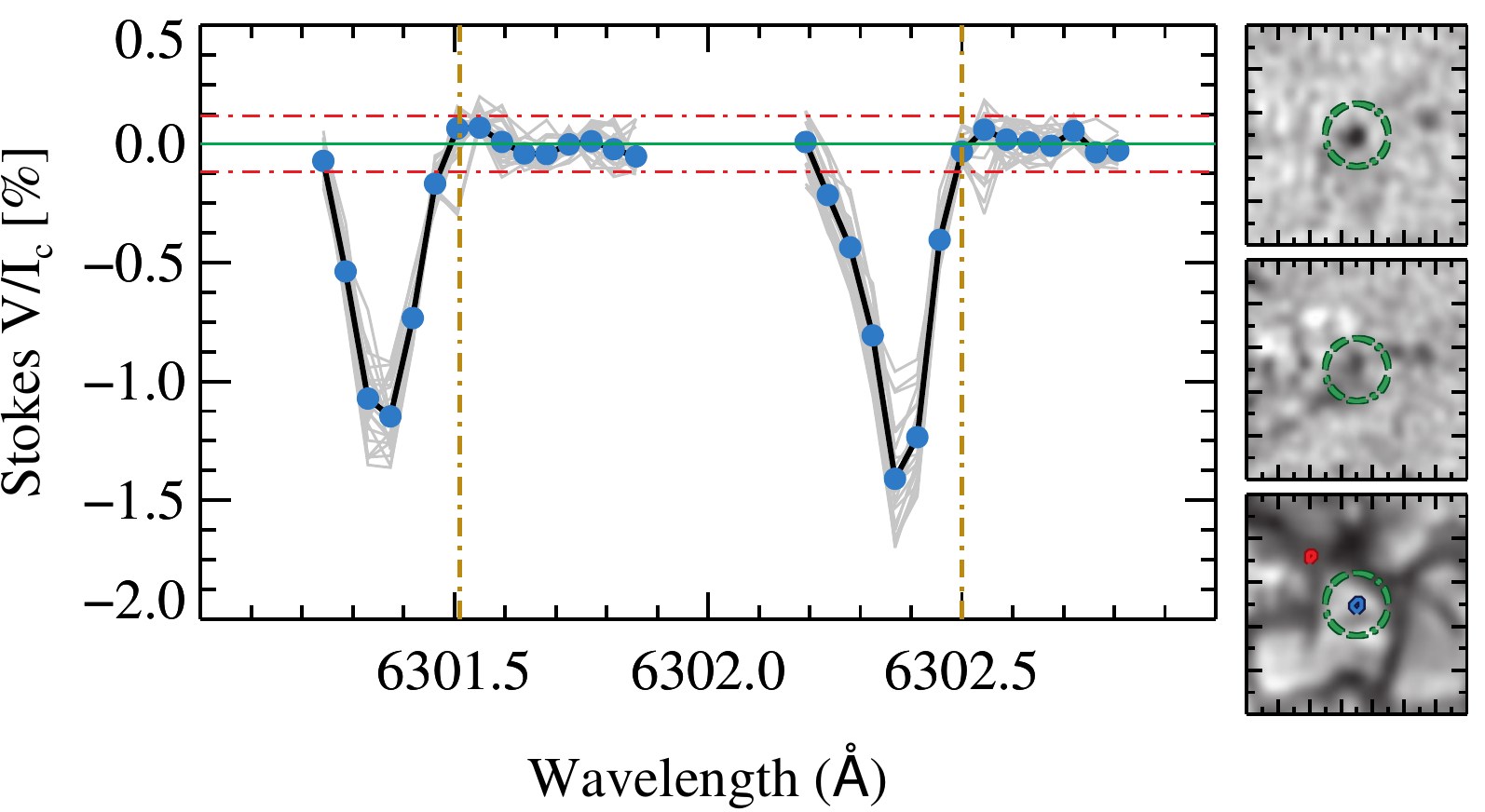}}\hfill
\renewcommand{\thesubfigure}{(III)}
  \subfloat[]{%
    \includegraphics[width=8.6cm, trim = 0 0 0 0, clip]{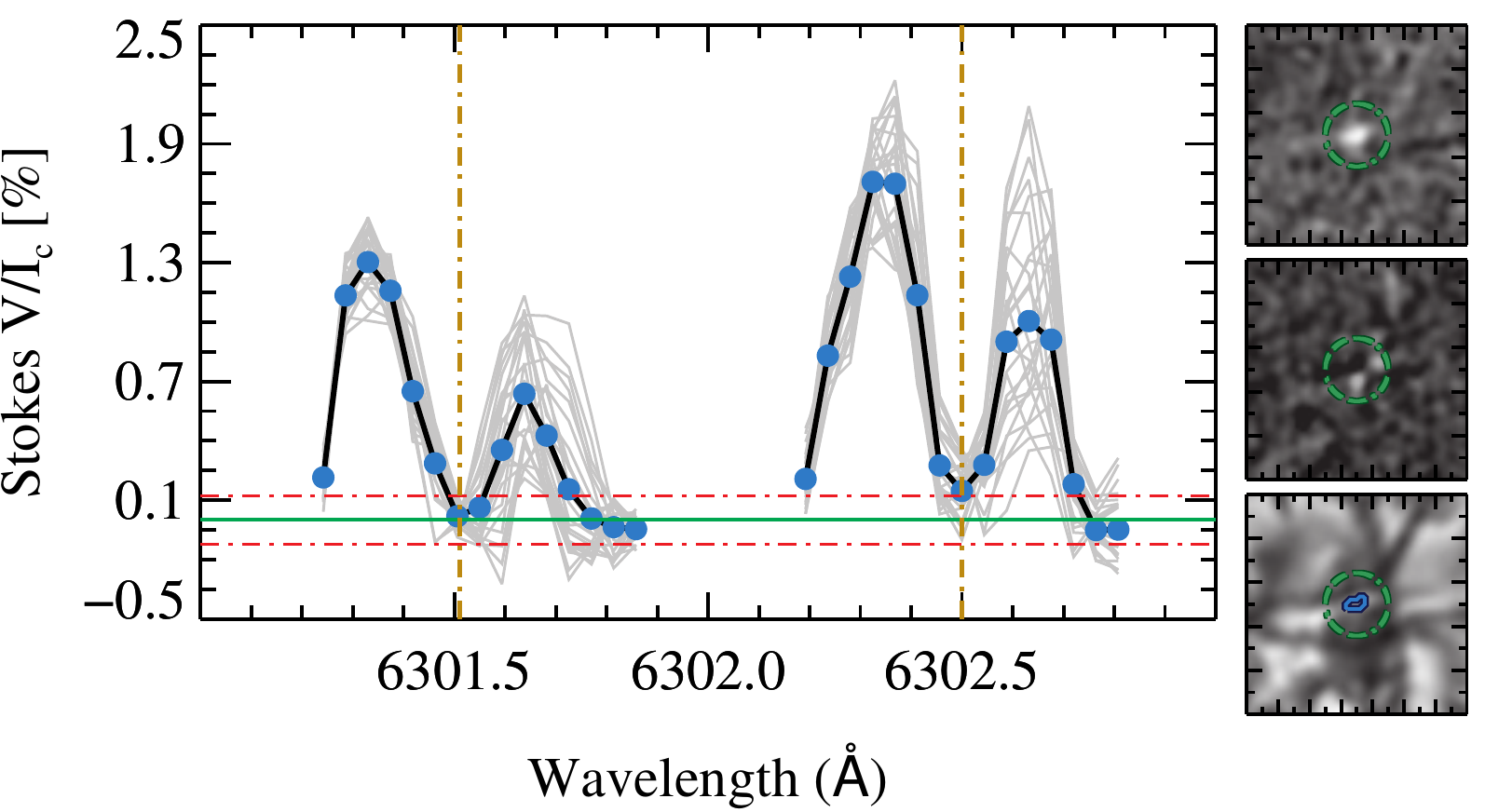}}\\
\renewcommand{\thesubfigure}{(IV)}
  \subfloat[]{%
    \includegraphics[width=8.6cm, trim = 0 0 0 0, clip]{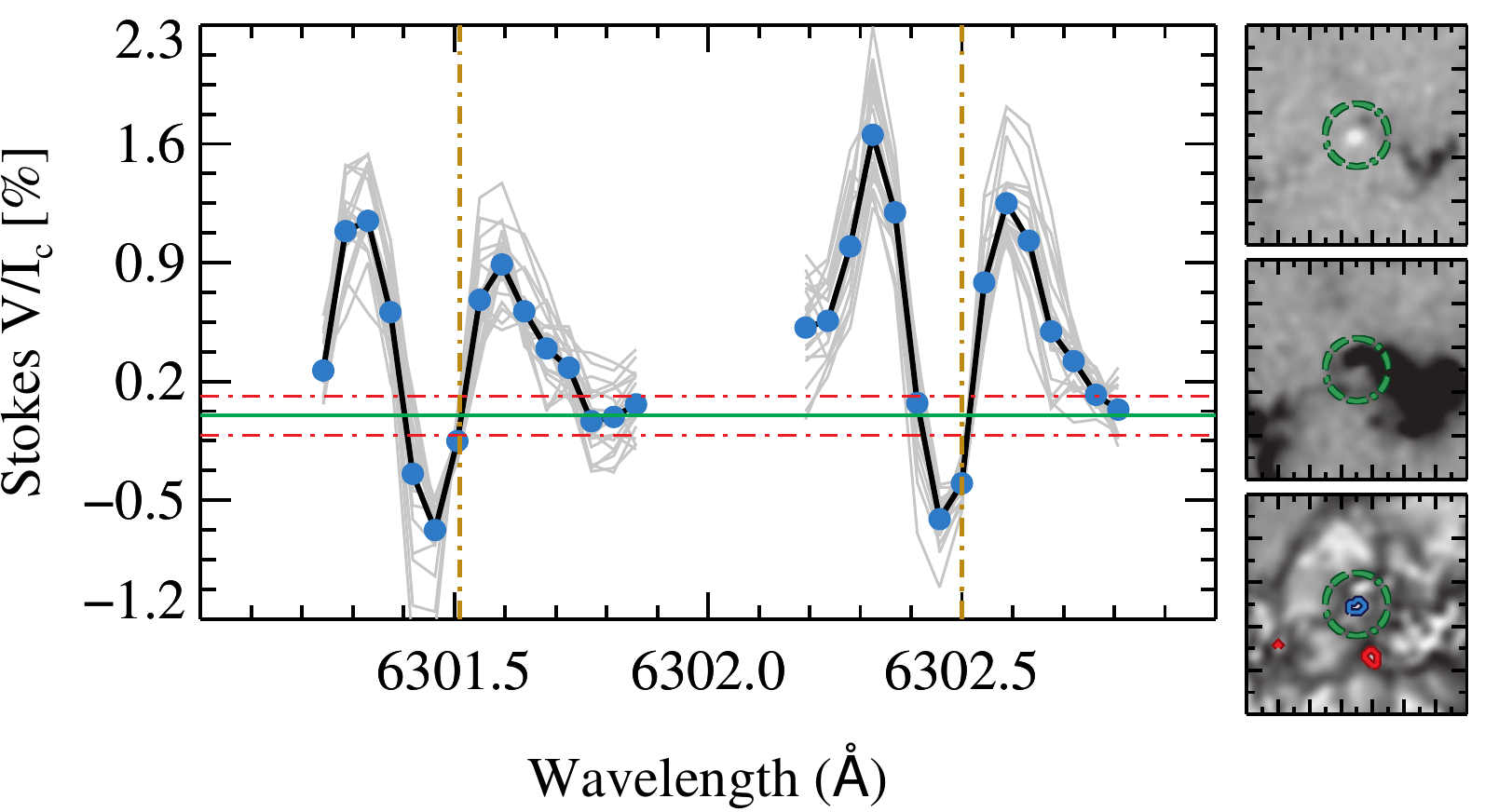}}\hfill
\renewcommand{\thesubfigure}{Reference (non-MUE)}
  \subfloat[]{%
    \includegraphics[width=8.6cm, trim = 0 0 0 0, clip]{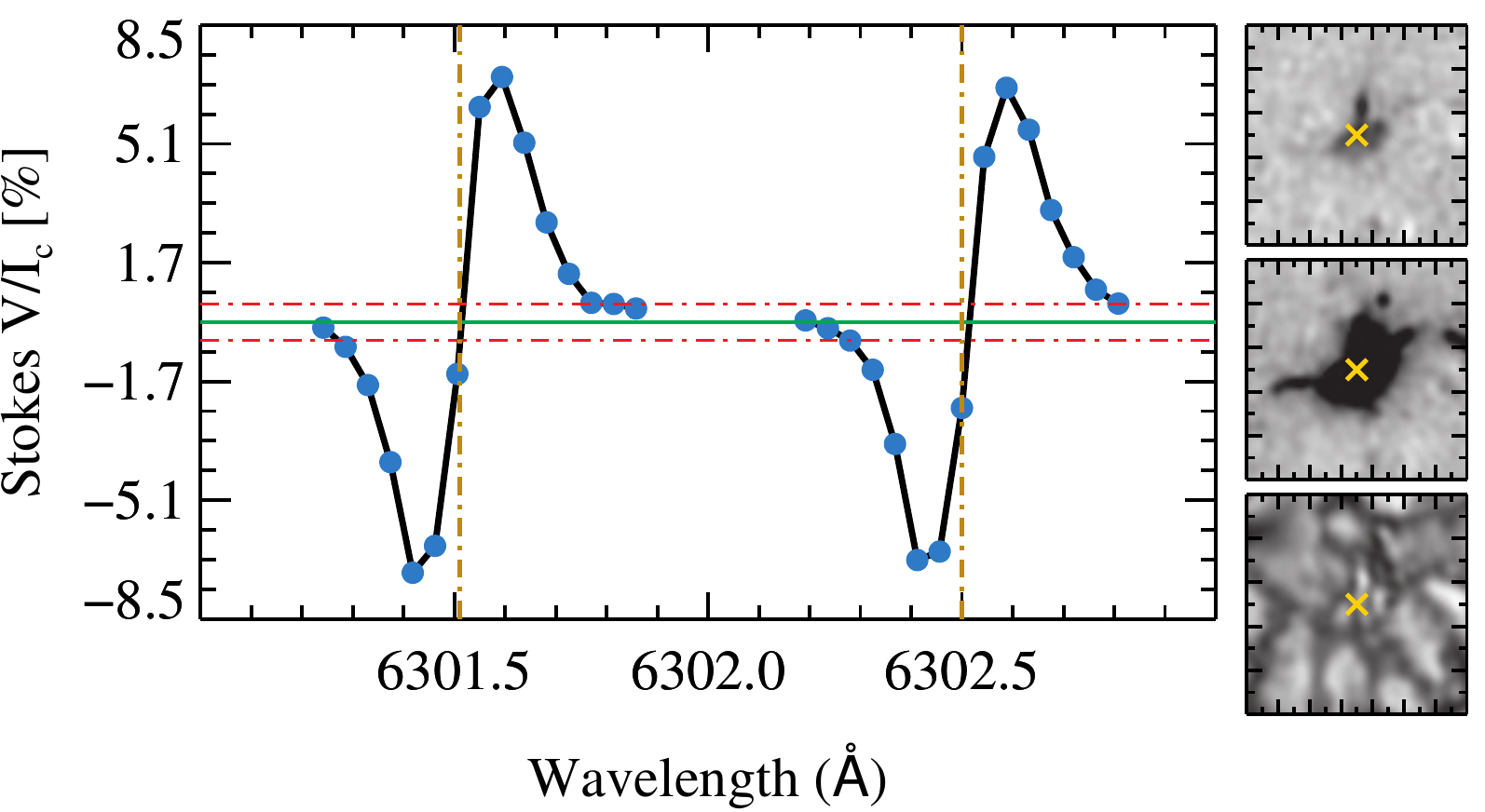}}\\
  \caption{Examples of different shapes (classes) of Stokes $V$ profiles of the magnetic upflow events (MUEs): (Ia) asymmetric but clear double-lobed profile, (Ib) asymmetric with a weak second lobe, (II) largely asymmetric, single-lobed profile, (III) a profile with two same-polarity lobes, and (IV) double-lobed profile with an extra blueshifted bump. The bottom right panel represents a typical example of a normal (double-lobe, antisymmetric) profile of a non-MUE as a reference. Plotted in gray are the profiles corresponding to all individual pixels of the MUEs marked by the green circles on the right panels. The mean profiles (black lines) as  well as the sampled wavelength positions (blue filled circles) are overlaid. The vertical lines mark the rest wavelengths of line cores of the 6301 and 6302\AA\ spectra. The red dash-dotted lines indicate the $3\sigma$ noise level of the corresponding averaged profiles. The right panels to the profiles show the images of far-blue-wing Stokes $V$ (top), in-line Stokes $V$ (middle), and continuum Stokes $I$ (bottom), with side lengths of $\approx2$~arcsec. The in-line $V$ images are saturated to the same scale as their corresponding far-blue-wing $V$ images. The blue and red contours in the continuum images indicate the locations of MUEs and MDEs in the field-of-view, respectively. The yellow cross in the bottom-right panel mark a network pixel for which the reference profile is plotted.}
  \label{fig:vprofiles}
\end{figure*}

Figure~\ref{fig:vprofiles} illustrates examples of the different classes of the observed profiles along with a typical example of a normal (reference) profile (bottom right). The latter represents the $V$ profile of a network pixel. In each sub-figure, three images are also plotted: the far-blue-wing Stokes $V$ (top), the in-line Stokes $V$ (middle), and Stokes $I$ continuum (bottom). In all three images, the green dotted-dashed circles mark the detected MUE. For comparisons, all the MUEs (including the one included in the green circles) and MDEs are outlined by blue and red contours on the continuum image. The Stokes $V$ profiles of all pixels within the detected MUE are plotted in gray in each panel. The mean $V$ profiles are overlaid in black along with their sampled wavelength positions (blue filled circles). The vertical lines mark the rest wavelengths of line core of the 6301 and 6302 \AA\ spectra.

Among all $V$ profiles of the $6301/2$~\AA\ pair lines, the fraction of different classes is (I) 27\%, (II) 42\%, (III) 23\%, and (IV) 3\%. The remaining profiles ($\approx5$\%) were excluded from our classification because of their peculiar or unclear shapes (i.e., they showed more complex appearances). These fractions are 26\%, 39\%, 17\%, and 5\% for classes (I) to (IV), respectively, from the $V$ profiles of the $5250$~\AA\ line. In addition, about 1\% of the 5250~\AA\ Stokes $V$ profiles appeared to be nearly normal (perfectly antisymmetric). Twelve percent of the profiles of the latter line falls into the unclear cases.

\section{Discussion and Conclusions}
\label{sec:conclusions}

We have presented a thorough observational study of MUEs from full Stokes observations of four photospheric magnetically sensitive lines recorded at high spatial, temporal, and spectral resolutions with the SST/CRISP instrument. We have proposed a robust definition from which the MUEs have been detected. The physical and dynamical properties of the MUEs have been provided and their Stokes $V$ profiles have been classified.

Our detection method (based on the difference between absolute values of significant Stokes $V$ signals in far-blue-wing and those in in-line wavelength positions; Section~\ref{subsec:subtract}), together with the high quality and largely sampled profiles, revealed a larger number of MUEs at any given time (larger by one to two orders of magnitude; $2.0\times10^{-2}$~arcsec$^{-2}$ MUEs covering $\approx0.25$\% of the entire quiet-Sun area, on average) compared to those reported in earlier studies ($\approx2.0\times10^{-3}$~arcsec$^{-2}$ in~\citealt{Borrero2010} and \citealt{Martinez-Pillet2011b}; $\approx7.0-9.0\times10^{-4}$~arcsec$^{-2}$ in~\citealt{Borrero2013} and \citealt{RubiodaCosta2015}). Some of the previous works were, however, limited by, e.g., few wavelength positions in their observed spectra and/or they were particularly aimed for enormous blueshifts in Stokes $V$ signals. These factors could result in a relatively small number of detected events. In the present work, by contrast, we aimed to study all Stokes $V$ signals which were blueshifted from the inner flanks of the studied lines. We note that our number of detected events was even influenced by our restrictive identification criteria with which we excluded MUEs smaller than 10 pixels in size and those with Stokes $V$ signals smaller than $7\sigma_\mathrm{noise}$. These have, however, reduced the chance of false detections.

\begin{figure*}[!t]
    \centerline{\includegraphics[width=18cm, trim = 0 0 0 0, clip]{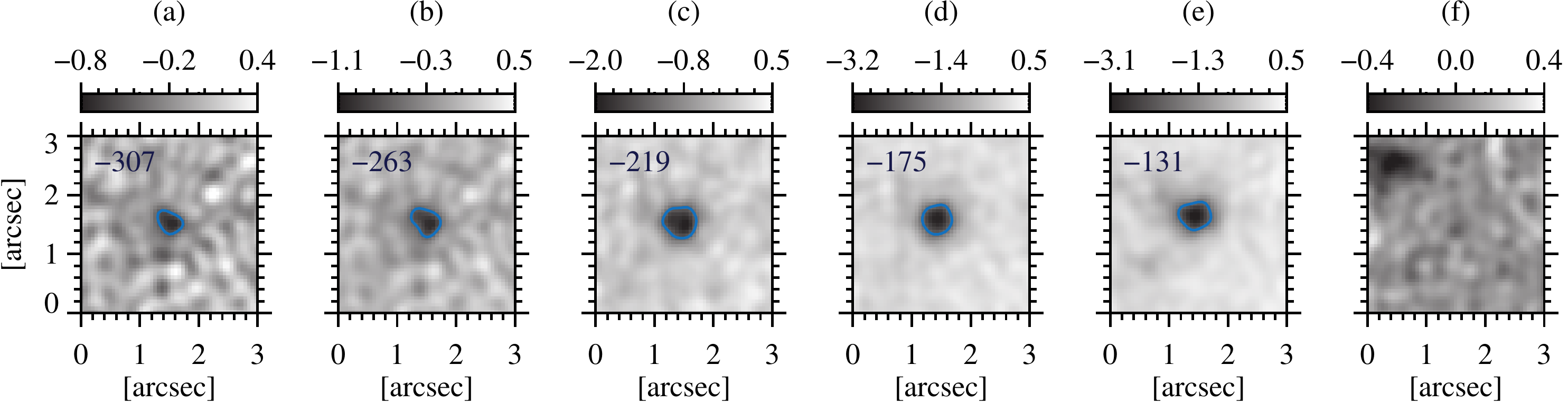}}
  \caption{One example of a magnetic upflow event (MUE) representing significant signals in all five far-blue-wing wavelength positions of a 6302~\AA\ Stokes $V$ profile (panels (a)-(e)). The numbers in the upper left corners of the panels (a)-(e) show the sampling wavelength positions in m\AA\ from the line core (the blue circles with minus signs in Figure~\ref{fig:stokesIps}(d)), which correspond to Doppler velocities of $-14.6$, $-12.5$, $-10.4$, $-8.3$, and $-6.2$~km$\:$s$^{-1}$, respectively. The blue contours outline the event where S/N~$\geq4$. The corresponding in-line Stokes $V$ image is shown in panel (f). All color bars show the Stokes $V$ signals in percentage, normalized to the Stokes $I$ continuum.}
  \label{fig:exampleMUE}
\end{figure*}

\citet{Borrero2010} found their events by identifying significant Stokes $V$ signals (signals larger than $\approx4\sigma_\mathrm{noise}$), in a single continuum position at $-227$~m\AA\ from the line core of Fe~{\sc i}~5250.653~\AA.\ Any significant Stokes $V$ signal was interpreted to stem from strongly blueshifted Fe~{\sc i}~5250.653~\AA\ components. The corresponding Doppler speed of 12~km$\:$s$^{-1}$ (to the wavelength difference of 227 m\AA)\ lead \citet{Borrero2010} to claim discovery of supersonic MUEs.

Presented in Figure~\ref{fig:exampleMUE} is an example where we find significant signals (signals larger than $4\sigma_\mathrm{noise}\approx0.64\%~I_{c}$) in all five wavelength positions in the far blue wing of 6302~\AA\ line (panels (a)-(e)). Panel (f) is the averaged in-line positions, given as the reference (where no signal is presented at the spatial location of the MUE). The five wavelength positions (the same as those marked by blue circles with minus signs in Figure~\ref{fig:stokesIps}(d)) were sampled in $-307$, $-263$, $-219$, $-175$, and $-131$ m\AA\ from the line core. These correspond to Doppler velocities of $-14.6$, $-12.5$, $-10.4$, $-8.3$, and $-6.2$~km$\:$s$^{-1}$, respectively. Thus, one could conclude that any of these rapid (supersonic) velocities is associated with the blueshifted Stokes $V$ signals, depending on the availability of these wavelength positions in the employed data. This may consequently result 
in a doubtful interpretation of Stokes $V$ signals at a single wavelength positions in the far blue wing in terms of supersonic upflow events. We note that this event is associated with a bisector LOS velocity of -2.1~km$\:$s$^{-1}$. We also note that all the detected MUEs do not show signals in all single wavelength positions of the far blue wings, but in some cases signals only exist in the bluest points whereas in other cases they exist only in the reddest positions. In any case, there is significant signal in, at least one of the five wavelength positions (or similarly for the other lines). Hence, the integrated far wings also provides the advantage of detecting a larger number of events compared to an approach based on one wavelength position in the far wings (for event detection).

Although \citet{Borrero2013} also formed in-line and far-blue-wing images from their coarsely sampled profiles, their $|V^{-}_\mathrm{wing}|$ actually represented very far blue-wing (or continuum-only) positions. Hence, that lead them to find the smallest number of events among the other studies (122 events, covering 857 pixels in a total area of $1.3\times10^{5}$~arcsec$^{2}$). However, we note that their study was aimed at only searching for highly blueshifted $V$ signals.
Here, we formed images with a relatively high S/N by averaging several wavelength positions at the wings and about the cores of the sampled lines prior to event detection. Hence, even with a more conservative signal threshold of $7\sigma_\mathrm{noise}$ of the treated images, we found a larger number of events also using the quotient method.

In order to investigate the influence of detection methods on the number of identified events, we also employed the quotient approach of \citet{Borrero2013} (quotient of absolute values of the far-blue-wing and in-line signals; Section~\ref{subsec:quotient}) on the same data sets used in the present work. Implementing the quotient method using the same criteria as in our difference approach (i.e., $|V^{-}_\mathrm{wing}|\geq7\sigma_\mathrm{noise}$ and size$>10$~pixels) results in a smaller number of MUEs, by a factor of $\approx2.5$ on average, compared to those we found earlier. With the same criterion as in \citet{Borrero2013} (i.e., when $|V^{-}_\mathrm{wing}/V_\mathrm{line}|>4$), we find thousands of pixels meeting this single condition, where no signal threshold is considered prior to or after the detection. The reason for this is that in the absence of a signal threshold, pixels with very weak signals (comparable with the noise level) could also satisfy $|V^{-}_\mathrm{wing}/V_\mathrm{line}|>4$.

The largest MUEs cover an area of about 0.45~Mm$^2$ ($\approx1$~~arcsec$^{2}$), while they are on average 0.06~Mm$^2$ in size. These are larger by a factor of two on average, compared to those found by \citet{Borrero2010} and \citet{RubiodaCosta2015}.
The Stokes $V$ signals of the MUEs range between $\approx0.6-4.0\times10^{-2}~I_{c}$ for all the passbands (reaching to a maximum signal of $4.9\times10^{-2}~I_{c}$ in some extreme cases) with a mean value of $1.3\times10^{-2}~I_{c}$. 
The MUEs live, on average, for $\approx110$~s, however, their lifetime distributions are extended toward 10~minutes (and a little bit beyond for a small fraction of the events). This is closely in agreement with the mean values of 81~s and 86~s reported by \citet{Borrero2010} and \citet{RubiodaCosta2015}, respectively.
The LOS velocities of our detected MUEs have a mean value of $-0.7$ to $-1.1$~km$\:$s$^{-1}$ that is consistent with those from \citet{RubiodaCosta2015}.
The average of sizes, Stokes $V$ signals, LOS velocities, and lifetimes of the MUEs are almost identical in observations in different wavelengths (the lifetimes were not determined for 5250~\AA\ MUEs because of the relatively short image-sequences of those observations). 
The 6173~\AA\ MUEs appeared to have a larger proper motion (mean value of 2.6~km$\:$s$^{-1}$) compared to those detected in the 6301 and 6302~\AA\ images (moving with a mean speed of 2.1, and 1.5~km$\:$s$^{-1}$, respectively). Those from 6173 and 6301~\AA\ tend to have relatively wide distributions of horizontal velocity extending toward the sound speed in the photosphere. This is not surprising, though, since these events are found to be mostly located at the edge of granules where rapid (sometimes supersonic) horizontal convective flows exist~\citep[e.g.,][]{Cattaneo1989,Solanki1996a,Rybak2004,Bellot2009,Nordlund2009,Vitas2011}.

A larger number of MUEs (by a factor of two) were also identified in 6173 \AA\ far-blue-wing images compared to those found in the other passbands.

The differences between the properties of the MUEs observed in the different passbands as well as the wide distributions of their physical quantities could depend on the intrinsic strength of the lines and/or on their excitation energies. These result in different heights of formation and in different temperature sensitivities of the lines. Thus, the variation of the physical parameters of the MUEs could be dependent on heights where they are placed. These differences could also be due to, e.g., different conditions of the different data sets observed in different times and at different regions (such as data quality, S/N, and level of solar activity).

MUEs smaller than 10 pixels in area and those with $V$ signals lower than $7\sigma_\mathrm{noise}$ were excluded in this study. These events may have different physical and dynamical properties. Hence, their actual distributions could be biased by these selection criteria which have, however, secured our results from the influence of falsely detected unresolved and/or relatively weak events.

We found about the same fraction of MUEs in network and internetwork areas. The spatial locations of the MUEs did not show any clear correlation with those of MDEs. There are, however, individual cases where an MUE is apparently located in the immediate vicinity of an MDE.

Furthermore, we classified Stokes $V$ profiles associated with the detected MUEs based on their appearances. We found them to fall into four classes: asymmetric double-lobed profiles (class I), asymmetric single-lobed profiles (class II), double-humped profiles (class III), and double-lobed profiles with an extra bump (class IV). The latter had an extra blueshifted bump in addition to a double-lobed profile (similar to those in class I), visually comparable with the appearance of Stokes $Q$ or $U$ profiles.

Different scenarios have been proposed for the physical processes involved in producing MUEs associated with single-lobed $V$ profiles. By inverting a larger spectrally sampled line of IMaX Fe~{\sc i} $5250$~\AA\ (compared to that of \citealt{Borrero2010}), \citet{Borrero2013} found coexistence of blue- and redshifted flows in about half of their detected events. From that, and also from observations of inclined magnetic fields of opposite polarities in the immediate vicinity of the events, they concluded that magnetic reconnection could be a possible physical mechanism for producing the extremely blueshifted magnetized flows. The latter process was confirmed by \citet{QuinteroNoda2013} who showed that interactions between emerging granular-scale loops and pre-existing fields in intergranular regions of {\sc Sunrise}/IMaX data could produce the strong upflows detected earlier. Later, \citet{QuinteroNoda2014} found that their detected MUEs from Hinode/SP images were often accompanied by MDEs, inferring structures corresponding to $\Omega$-loops. They concluded that siphon flows along arched, magnetic flux tubes could explain those field configurations (including the magnetic flows along them).

MHD simulations of emergence and cancellation of magnetic flux in the solar photosphere led \citet{Danilovic2015} to conclude that magnetic flux emergence is enough to produce the rapid magnetic flows where magnetic reconnection may occur but not necessarily.

Note that all above mechanisms discuss the magnetic upflows corresponding to single-lobed Stokes $V$ profiles. Different physical processes may be involved in producing MUEs with different shapes of Stokes $V$ profiles (as we found in the present work). 

The highly asymmetric profiles (e.g., the single-lobed profiles; the class II) are, in general, interpreted as the results of strong gradients in the LOS velocity and/or the vector magnetic field (e.g., \citealt{Illing1975,Landi-DeglInnocenti1983,Grossmann-Doerth2000}; see also \citealt{SainzDalda2012}). This may also include the profiles in classes Ia and Ib whose second lobes have stronger signals compared to those in class II.
The level of observed asymmetry in $V$ profiles, as a result of velocity gradients, has been shown to be dependent on the strength (magnetic sensitivity) of the line, i.e., the velocity gradient produces almost no asymmetry in weak lines~\citep{Solanki1988}. This may also, however, depend on the S/N of the observations.
In addition to the line strength, the dependence of the asymmetries have been described on Zeeman splitting, Doppler shift, and line width~\citep{Grossmann-Doerth1989}. The dependency of the line asymmetry on the line strength and the Doppler shift could possibly explain the somewhat different fractions of asymmetric $V$ profiles we found from the 5250 and the 6301/2 \AA\ lines.

The double-humped profiles, class (III), as well as the double-lobed profiles with an extra blueshifted bump, class (IV), could be examples of mixed polarity Stokes $V$ profiles. In this case, these profiles could be produced as a result of, e.g., a spatially unresolved mixture of opposite polarities~\citep{Grossmann-Doerth2000,Sigwarth2001}. \citet{Socas-Navarro2005} found a few supersonic MUEs with three-lobed Stokes $V$ profiles, similar to those that fell into our class (IV). They described the extra blueshifted bump as a possible signature of an exploding magnetic element due to aborted convective collapse~\citep{BellotRubio2001}. The two classes of Stokes $V$ profiles (i.e., the double-humped and the double-lobed with an extra bump) could, however, be produced from a region of unipolar magnetic field (using a two-layer model atmosphere) by \citet{Steiner2000}, who also provided a wide variety of Stokes $V$ profiles. They showed that while one lobe of the double-humped profiles (as a particular case of $Q$-like Stokes $V$ profiles) was in absorption, the other lobe could be in partial or complete emission. The latter was explained as a result of temperature inversion where the temperature of the magnetic layer surpassed the temperature of the line-core-forming regions. Thus, \citet{Steiner2000} showed that the line depth plays an important role in forming these types of $V$ profiles.

We note that large blueshifted signals in Stokes $V$ profiles do not imply the presence of large (supersonic) velocities from our bisector analysis; even so, they cannot be completely ruled out either. In general, the largely Doppler-shifted signals could be due to, e.g., high temperatures that could increase the Doppler width, a large line shift as a result of strong flows, a very high pressure causing strong broadening of the line, or a turbulence velocity making an additional broadening.
Therefore, at least for a part of the flows, the large blueshifted signals could correspond to very large velocities. Determining these large velocities may, however, not be straightforward. The latter needs careful interpretations of the profiles, e.g., from Stokes inversions and/or from MHD simulations to distinguish the possible thermal widths from the fast flows.

To summarize, we found a larger number of MUEs that was larger by one to two orders of magnitude than previously reported. This has been shown to be dependent on detection approach, on S/N, and possibly on wavelength and resolutions of the employed data sets. The MUEs happened to be approximately equally distributed in network and internetwork areas. A larger number of events (by a factor of $\approx2$) were identified in 6173 \AA\ data sets compared to those observed in 5250 \AA\ and 6301/2 \AA. The MUEs detected in the former images tend to move faster (by a factor of 1.7 on average) in the horizontal plane than those found in 6302 \AA\ spectra. The MUEs represent different shapes of Stokes $V$ profiles of which only less than half of them are single-lobed. The rest of the profiles appear to be either asymmetric double-lobed or illustrate signatures of mixed polarities. Some debates also exist on the origins and on the natures of MUEs (of different kinds). Hence, their tentative interpretations need further investigations from, e.g., detailed Stokes inversions and/or realistic MHD simulations of various magnetically sensitive lines. It also includes a study of their structures and temporal evolutions, characterization of the different classes of $V$ profiles associated with MUEs, and chasing them in the higher atmospheric layers of the Sun.

\acknowledgements
S.J. is grateful to J.M.~Borerro, S.K.~Solanki, and M.~van~Noort for useful discussions. We thank A.~Cristaldi, M.~Falco, and S.~Criscuoli for SST observations and for contributions to the data processing of the 6301/2 \AA\ data sets. The Swedish 1 m Solar Telescope is operated on the island of La Palma by the Institute for Solar Physics of Stockholm University in the Spanish Observatorio del Roque de los Muchachos of the Instituto de Astrof\' isica de Canarias.

\bibliographystyle{aa}
\bibliography{ApJ98854}

\begin{thebibliography}{63}
\expandafter\ifx\csname natexlab\endcsname\relax\def\natexlab#1{#1}\fi

\bibitem[{{Auer} {et~al.}(1977){Auer}, {House}, \& {Heasley}}]{Auer1977}
{Auer}, L.~H., {House}, L.~L., \& {Heasley}, J.~N. 1977, SoPh, 55, 47

\bibitem[{{Bellot Rubio}(2009)}]{Bellot2009}
{Bellot Rubio}, L.~R. 2009, \apj, 700, 284

\bibitem[{{Bellot Rubio} {et~al.}(2001){Bellot Rubio}, {Rodr{\'{\i}}guez
  Hidalgo}, {Collados}, {Khomenko}, \& {Ruiz Cobo}}]{BellotRubio2001}
{Bellot Rubio}, L.~R., {Rodr{\'{\i}}guez Hidalgo}, I., {Collados}, M.,
  {Khomenko}, E., \& {Ruiz Cobo}, B. 2001, \apj, 560, 1010

\bibitem[{{Borrero} \& {Ichimoto}(2011)}]{Borrero2011}
{Borrero}, J.~M. \& {Ichimoto}, K. 2011, LRSP, 8, 4

\bibitem[{{Borrero} {et~al.}(2010){Borrero}, {Mart{\'{\i}}nez-Pillet},
  {Schlichenmaier}, {Solanki}, {Bonet}, {del Toro Iniesta}, {Schmidt},
  {Barthol}, {Gandorfer}, {Domingo}, \& {Kn{\"o}lker}}]{Borrero2010}
{Borrero}, J.~M., {Mart{\'{\i}}nez-Pillet}, V., {Schlichenmaier}, R., {et~al.}
  2010, ApJL, 723, L144

\bibitem[{{Borrero} {et~al.}(2013){Borrero}, {Mart{\'{\i}}nez-Pillet},
  {Schmidt}, {Quintero Noda}, {Bonet}, {del Toro Iniesta}, \& {Bellot
  Rubio}}]{Borrero2013}
{Borrero}, J.~M., {Mart{\'{\i}}nez-Pillet}, V., {Schmidt}, W., {et~al.} 2013,
  \apj, 768, 69

\bibitem[{Caligari {et~al.}(1995)Caligari, Moreno-Insertis, \&
  Sch{\"{u}}ssler}]{Caligari1995}
Caligari, P., Moreno-Insertis, F., \& Sch{\"{u}}ssler, M. 1995, \apj, 441, 886

\bibitem[{{Cattaneo} {et~al.}(1989){Cattaneo}, {Hurlburt}, \&
  {Toomre}}]{Cattaneo1989}
{Cattaneo}, F., {Hurlburt}, N.~E., \& {Toomre}, J. 1989, in NATO ASIC Proc.
  263: Solar and Stellar Granulation, ed. {R.~J.~{Rutten} \& and G.~{Severino}
  (Dordrecht: Kluwer)}, 415

\bibitem[{{Danilovic} {et~al.}(2010){Danilovic}, {Beeck}, {Pietarila},
  {Sch{\"u}ssler}, {Solanki}, {Mart{\'{\i}}nez-Pillet}, {Bonet}, {del Toro
  Iniesta}, {Domingo}, {Barthol}, {Berkefeld}, {Gandorfer}, {Kn{\"o}lker},
  {Schmidt}, \& {Title}}]{Danilovic2010b}
{Danilovic}, S., {Beeck}, B., {Pietarila}, A., {et~al.} 2010, ApJL, 723, L149

\bibitem[{{Danilovic} {et~al.}(2015){Danilovic}, {Cameron}, \&
  {Solanki}}]{Danilovic2015}
{Danilovic}, S., {Cameron}, R.~H., \& {Solanki}, S.~K. 2015, \aap, 574, A28

\bibitem[{{de la Cruz Rodr{\'{\i}}guez} {et~al.}(2015){de la Cruz
  Rodr{\'{\i}}guez}, {L{\"o}fdahl}, {S{\"u}tterlin}, {Hillberg}, \& {Rouppe van
  der Voort}}]{de-laCruz2015}
{de la Cruz Rodr{\'{\i}}guez}, J., {L{\"o}fdahl}, M.~G., {S{\"u}tterlin}, P.,
  {Hillberg}, T., \& {Rouppe van der Voort}, L. 2015, \aap, 573, A40

\bibitem[{{de Wijn} {et~al.}(2009){de Wijn}, {Stenflo}, {Solanki}, \&
  {Tsuneta}}]{deWijn2009}
{de Wijn}, A.~G., {Stenflo}, J.~O., {Solanki}, S.~K., \& {Tsuneta}, S. 2009,
  SSRv, 144, 275

\bibitem[{{Fabbian} {et~al.}(2012){Fabbian}, {Moreno-Insertis}, {Khomenko}, \&
  {Nordlund}}]{Fabbian2012}
{Fabbian}, D., {Moreno-Insertis}, F., {Khomenko}, E., \& {Nordlund}, {\AA}.
  2012, \aap, 548, A35

\bibitem[{{Fischer} {et~al.}(2009){Fischer}, {de Wijn}, {Centeno}, {Lites}, \&
  {Keller}}]{Fischer2009}
{Fischer}, C.~E., {de Wijn}, A.~G., {Centeno}, R., {Lites}, B.~W., \& {Keller},
  C.~U. 2009, \aap, 504, 583

\bibitem[{{Fontenla} {et~al.}(2006){Fontenla}, {Avrett}, {Thuillier}, \&
  {Harder}}]{Fontenla2006}
{Fontenla}, J.~M., {Avrett}, E., {Thuillier}, G., \& {Harder}, J. 2006, \apj,
  639, 441

\bibitem[{{Fontenla} {et~al.}(1993){Fontenla}, {Avrett}, \&
  {Loeser}}]{Fontenla1993}
{Fontenla}, J.~M., {Avrett}, E.~H., \& {Loeser}, R. 1993, \apj, 406, 319

\bibitem[{{Grossmann-Doerth} {et~al.}(1989){Grossmann-Doerth}, {Schuessler}, \&
  {Solanki}}]{Grossmann-Doerth1989}
{Grossmann-Doerth}, U., {Schuessler}, M., \& {Solanki}, S.~K. 1989, \aap, 221,
  338

\bibitem[{{Grossmann-Doerth} {et~al.}(2000){Grossmann-Doerth}, {Sch{\"u}ssler},
  {Sigwarth}, \& {Steiner}}]{Grossmann-Doerth2000}
{Grossmann-Doerth}, U., {Sch{\"u}ssler}, M., {Sigwarth}, M., \& {Steiner}, O.
  2000, \aap, 357, 351

\bibitem[{{Illing} {et~al.}(1975){Illing}, {Landman}, \& {Mickey}}]{Illing1975}
{Illing}, R.~M.~E., {Landman}, D.~A., \& {Mickey}, D.~L. 1975, \aap, 41, 183

\bibitem[{{Jafarzadeh} {et~al.}(2013){Jafarzadeh}, {Solanki}, {Feller}, {Lagg},
  {Pietarila}, {Danilovic}, {Riethm{\"u}ller}, \& {Mart{\'{\i}}nez
  Pillet}}]{Jafarzadeh2013a}
{Jafarzadeh}, S., {Solanki}, S.~K., {Feller}, A., {et~al.} 2013, \aap, 549,
  A116

\bibitem[{Kramida {et~al.}(2014)Kramida, {Yu.~Ralchenko}, Reader, \& {NIST ASD
  Team}}]{Kramida2014}
Kramida, A., {Yu.~Ralchenko}, Reader, J., \& {NIST ASD Team}. 2014, NIST
  database, {NIST Atomic Spectra Database (ver. 5.2), [Online]. Available:
  {\tt{http://physics.nist.gov/asd}} [2015, April 19]. National Institute of
  Standards and Technology, Gaithersburg, MD.}

\bibitem[{{Landi Degl'Innocenti} \& {Landolfi}(1983)}]{Landi-DeglInnocenti1983}
{Landi Degl'Innocenti}, E. \& {Landolfi}, M. 1983, SoPh, 87, 221

\bibitem[{{Landi Degl'Innocenti} \&
  {Landolfi}(2004)}]{Landi-DeglInnocenti12004}
{Landi Degl'Innocenti}, E. \& {Landolfi}, M., eds. 2004, {Polarization in
  Spectral Lines}, Vol. 307

\bibitem[{{Lites} {et~al.}(1996){Lites}, {Leka}, {Skumanich},
  {Mart{\'{\i}}nez-Pillet}, \& {Shimizu}}]{Lites1996}
{Lites}, B.~W., {Leka}, K.~D., {Skumanich}, A., {Mart{\'{\i}}nez-Pillet}, V.,
  \& {Shimizu}, T. 1996, \apj, 460, 1019

\bibitem[{{Mart{\'{\i}}nez Gonz{\'a}lez} \& {Bellot
  Rubio}(2009)}]{Martinez-Gonzalez2009}
{Mart{\'{\i}}nez Gonz{\'a}lez}, M.~J. \& {Bellot Rubio}, L.~R. 2009, \apj, 700,
  1391

\bibitem[{{Mart{\'{\i}}nez-Pillet} {et~al.}(2011){Mart{\'{\i}}nez-Pillet}, {del
  Toro Iniesta}, {{\'A}lvarez-Herrero}, {Domingo}, {Bonet}, {Gonz{\'a}lez
  Fern{\'a}ndez}, {L{\'o}pez Jim{\'e}nez}, {Pastor}, {Gasent Blesa}, {Mellado},
  {Piqueras}, {Aparicio}, {Balaguer}, {Ballesteros}, {Belenguer}, {Bellot
  Rubio}, {Berkefeld}, {Collados}, {Deutsch}, {Feller}, {Girela}, {Grauf},
  {Heredero}, {Herranz}, {Jer{\'o}nimo}, {Laguna}, {Meller}, {Men{\'e}ndez},
  {Morales}, {Orozco Su{\'a}rez}, {Ramos}, {Reina}, {Ramos},
  {Rodr{\'{\i}}guez}, {S{\'a}nchez}, {Uribe-Patarroyo}, {Barthol}, {Gandorfer},
  {Kn{\"o}lker}, {Schmidt}, {Solanki}, \& {Vargas
  Dom{\'{\i}}nguez}}]{Martinez-Pillet2011}
{Mart{\'{\i}}nez-Pillet}, V., {del Toro Iniesta}, J.~C., {{\'A}lvarez-Herrero},
  A., {et~al.} 2011, SoPh, 268, 57

\bibitem[{{Mart{\'{\i}}nez Pillet} {et~al.}(2011){Mart{\'{\i}}nez Pillet}, {del
  Toro Iniesta}, \& {Quintero Noda}}]{Martinez-Pillet2011b}
{Mart{\'{\i}}nez Pillet}, V., {del Toro Iniesta}, J.~C., \& {Quintero Noda}, C.
  2011, \aap, 530, A111

\bibitem[{{Montesinos} \& {Thomas}(1993)}]{Montesinos1993}
{Montesinos}, B. \& {Thomas}, J.~H. 1993, \apj, 402, 314

\bibitem[{{Nordlund} {et~al.}(2009){Nordlund}, {Stein}, \&
  {Asplund}}]{Nordlund2009}
{Nordlund}, {\AA}., {Stein}, R.~F., \& {Asplund}, M. 2009, LRSP, 6, 2. URL:
  http://www.livingreviews.org/lrsp

\bibitem[{{Orozco Su{\'a}rez} {et~al.}(2008){Orozco Su{\'a}rez}, {Bellot
  Rubio}, {del Toro Iniesta}, \& {Tsuneta}}]{Orozco2008}
{Orozco Su{\'a}rez}, D., {Bellot Rubio}, L.~R., {del Toro Iniesta}, J.~C., \&
  {Tsuneta}, S. 2008, \aap, 481, L33

\bibitem[{{Quintero Noda} {et~al.}(2014){Quintero Noda}, {Borrero}, {Orozco
  Su{\'a}rez}, \& {Ruiz Cobo}}]{QuinteroNoda2014}
{Quintero Noda}, C., {Borrero}, J.~M., {Orozco Su{\'a}rez}, D., \& {Ruiz Cobo},
  B. 2014, \aap, 569, A73

\bibitem[{{Quintero Noda} {et~al.}(2013){Quintero Noda},
  {Mart{\'{\i}}nez-Pillet}, {Borrero}, \& {Solanki}}]{QuinteroNoda2013}
{Quintero Noda}, C., {Mart{\'{\i}}nez-Pillet}, V., {Borrero}, J.~M., \&
  {Solanki}, S.~K. 2013, \aap, 558, A30

\bibitem[{{Rachkovsky}(1962)}]{Rachkovsky1962}
{Rachkovsky}, D.~N. 1962, Izv. Krymsk. Astrofiz. Obs, 28, 259

\bibitem[{{Rubio da Costa} {et~al.}(2015){Rubio da Costa}, {Solanki},
  {Danilovic}, {Hizberger}, \& {Mart{\'{\i}}nez-Pillet}}]{RubiodaCosta2015}
{Rubio da Costa}, F., {Solanki}, S.~K., {Danilovic}, S., {Hizberger}, J., \&
  {Mart{\'{\i}}nez-Pillet}, V. 2015, \aap, 574, A95

\bibitem[{{Ryb{\'a}k} {et~al.}(2004){Ryb{\'a}k}, {W{\"o}hl}, {Ku{\v c}era},
  {Hanslmeier}, \& {Steiner}}]{Rybak2004}
{Ryb{\'a}k}, J., {W{\"o}hl}, H., {Ku{\v c}era}, A., {Hanslmeier}, A., \&
  {Steiner}, O. 2004, \aap, 420, 1141

\bibitem[{{Sainz Dalda} {et~al.}(2012){Sainz Dalda}, {Mart{\'{\i}}nez-Sykora},
  {Bellot Rubio}, \& {Title}}]{SainzDalda2012}
{Sainz Dalda}, A., {Mart{\'{\i}}nez-Sykora}, J., {Bellot Rubio}, L., \&
  {Title}, A. 2012, \apj, 748, 38

\bibitem[{{Scharmer}(2006)}]{Scharmer2006}
{Scharmer}, G.~B. 2006, \aap, 447, 1111

\bibitem[{{Scharmer} {et~al.}(2003){Scharmer}, {Bjelksjo}, {Korhonen},
  {Lindberg}, \& {Petterson}}]{Scharmer2003}
{Scharmer}, G.~B., {Bjelksjo}, K., {Korhonen}, T.~K., {Lindberg}, B., \&
  {Petterson}, B. 2003, in , 341

\bibitem[{{Scharmer} {et~al.}(2008){Scharmer}, {Narayan}, {Hillberg}, {de la
  Cruz Rodr{\'{\i}}guez}, {L{\"o}fdahl}, {Kiselman}, {S{\"u}tterlin}, {van
  Noort}, \& {Lagg}}]{Scharmer2008}
{Scharmer}, G.~B., {Narayan}, G., {Hillberg}, T., {et~al.} 2008, ApJL, 689, L69

\bibitem[{{Shimizu} {et~al.}(2008){Shimizu}, {Lites}, {Katsukawa}, {Ichimoto},
  {Suematsu}, {Tsuneta}, {Nagata}, {Kubo}, {Shine}, \& {Tarbell}}]{Shimizu2008}
{Shimizu}, T., {Lites}, B.~W., {Katsukawa}, Y., {et~al.} 2008, \apj, 680, 1467

\bibitem[{{Shimizu} {et~al.}(2007){Shimizu}, {Martinez-Pillet}, {Collados},
  {Ruiz-Cobo}, {Centeno}, {Beck}, \& {Katsukawa}}]{Shimizu2007b}
{Shimizu}, T., {Martinez-Pillet}, V., {Collados}, M., {et~al.} 2007, in New
  Solar Physics with Solar-B Mission, ed. K.~{Shibata}, S.~{Nagata}, \&
  T.~{Sakurai}, Vol. 369 (San Francisco, CA: ASP), 113

\bibitem[{{Sigwarth}(2001)}]{Sigwarth2001}
{Sigwarth}, M. 2001, \apj, 563, 1031

\bibitem[{{Socas-Navarro}(2004)}]{Socas-Navarro2004}
{Socas-Navarro}, H. 2004, \apj, 613, 610

\bibitem[{{Socas-Navarro} \& {Manso Sainz}(2005)}]{Socas-Navarro2005}
{Socas-Navarro}, H. \& {Manso Sainz}, R. 2005, ApJL, 620, L71

\bibitem[{{Solanki}(1993)}]{Solanki1993}
{Solanki}, S.~K. 1993, \ssr, 63, 1

\bibitem[{{Solanki}(2001)}]{Solanki2001}
{Solanki}, S.~K. 2001, in Magnetic Fields Across the Hertzsprung-Russell
  Diagram, ed. {{Mathys}, G. and {Solanki}, S.~K. and {Wickramasinghe}, D.~T.},
  Vol. 248 (San Francisco, CA: ASP), 45

\bibitem[{{Solanki} {et~al.}(2010){Solanki}, {Barthol}, {Danilovic}, {Feller},
  {Gandorfer}, {Hirzberger}, {Riethm{\"u}ller}, {Sch{\"u}ssler}, {Bonet},
  {Mart{\'{\i}}nez Pillet}, {del Toro Iniesta}, {Domingo}, {Palacios},
  {Kn{\"o}lker}, {Bello Gonz{\'a}lez}, {Berkefeld}, {Franz}, {Schmidt}, \&
  {Title}}]{Solanki2010}
{Solanki}, S.~K., {Barthol}, P., {Danilovic}, S., {et~al.} 2010, ApJL, 723,
  L127

\bibitem[{{Solanki} \& {Pahlke}(1988)}]{Solanki1988}
{Solanki}, S.~K. \& {Pahlke}, K.~D. 1988, \aap, 201, 143

\bibitem[{{Solanki} {et~al.}(1996){Solanki}, {Rueedi}, {Bianda}, \&
  {Steffen}}]{Solanki1996a}
{Solanki}, S.~K., {Rueedi}, I., {Bianda}, M., \& {Steffen}, M. 1996, \aap, 308,
  623

\bibitem[{{Spruit} {et~al.}(1990){Spruit}, {Nordlund}, \& {Title}}]{Spruit1990}
{Spruit}, H.~C., {Nordlund}, A., \& {Title}, A.~M. 1990, \araa, 28, 263

\bibitem[{{Steiner}(2000)}]{Steiner2000}
{Steiner}, O. 2000, SoPh, 196, 245

\bibitem[{{Steiner} {et~al.}(2010){Steiner}, {Franz}, {Bello Gonz{\'a}lez},
  {Nutto}, {Rezaei}, {Mart{\'{\i}}nez Pillet}, {Bonet Navarro}, {del Toro
  Iniesta}, {Domingo}, {Solanki}, {Kn{\"o}lker}, {Schmidt}, {Barthol}, \&
  {Gandorfer}}]{Steiner2010}
{Steiner}, O., {Franz}, M., {Bello Gonz{\'a}lez}, N., {et~al.} 2010, ApJL, 723,
  L180

\bibitem[{{Stenflo}(1971)}]{Stenflo1971}
{Stenflo}, J.~O. 1971, in IAU Symp. (Dordrecht: Reidel), Vol.~43, Solar
  Magnetic Fields, ed. R.~{Howard}, 101

\bibitem[{{Stenflo}(1989)}]{Stenflo1989}
{Stenflo}, J.~O. 1989, A\&ARv, 1, 3

\bibitem[{{Straus} {et~al.}(2010){Straus}, {Fleck}, {Jefferies}, {Carlsson}, \&
  {Tarbell}}]{Straus2010}
{Straus}, T., {Fleck}, B., {Jefferies}, S.~M., {Carlsson}, M., \& {Tarbell},
  T.~D. 2010, MmSAI, 81, 751

\bibitem[{{Tsuneta} {et~al.}(2008){Tsuneta}, {Ichimoto}, {Katsukawa}, {Nagata},
  {Otsubo}, {Shimizu}, {Suematsu}, {Nakagiri}, {Noguchi}, {Tarbell}, {Title},
  {Shine}, {Rosenberg}, {Hoffmann}, {Jurcevich}, {Kushner}, {Levay}, {Lites},
  {Elmore}, {Matsushita}, {Kawaguchi}, {Saito}, {Mikami}, {Hill}, \&
  {Owens}}]{Tsuneta2008}
{Tsuneta}, S., {Ichimoto}, K., {Katsukawa}, Y., {et~al.} 2008, SoPh, 249, 167

\bibitem[{{Uitenbroek}(2001)}]{Uitenbroek2001}
{Uitenbroek}, H. 2001, \apj, 557, 389

\bibitem[{{Unno}(1956)}]{Unno1956}
{Unno}, W. 1956, \pasj, 8, 108

\bibitem[{{van Noort} {et~al.}(2005){van Noort}, {Rouppe van der Voort}, \&
  {L{\"o}fdahl}}]{vanNoort2005}
{van Noort}, M., {Rouppe van der Voort}, L., \& {L{\"o}fdahl}, M.~G. 2005,
  SoPh, 228, 191

\bibitem[{{van Noort} \& {Rouppe van der Voort}(2008)}]{vanNoort2008}
{van Noort}, M.~J. \& {Rouppe van der Voort}, L.~H.~M. 2008, \aap, 489, 429

\bibitem[{{Vitas} {et~al.}(2011){Vitas}, {Fischer}, {V{\"o}gler}, \&
  {Keller}}]{Vitas2011}
{Vitas}, N., {Fischer}, C.~E., {V{\"o}gler}, A., \& {Keller}, C.~U. 2011, \aap,
  532, A110

\bibitem[{{Wittmann}(1974)}]{Wittmann1974}
{Wittmann}, A. 1974, SoPh, 35, 11

\bibitem[{{Zhang} {et~al.}(2009){Zhang}, {Yang}, \& {Jin}}]{Zhang2009}
{Zhang}, J., {Yang}, S.-H., \& {Jin}, C.-L. 2009, RAA, 9, 921

\end{thebibliography}

\end{document}